\documentclass{aastex63}

\received{August, 2020}
\revised{December, 2020}
\accepted{December, 2020}
\shorttitle{GRB central engine}
\shortauthors{Sharma V. et al.}
\graphicspath{{./}{figures/}}

\begin{document}

\title{Identifying Black Hole Central Engines in Gamma-Ray Bursts}

\correspondingauthor{Vidushi Sharma, Shabnam Iyyani}
\email{vidushi@iucaa.in, shabnam@iucaa.in}

\author[0000-0002-0786-7307]{Vidushi Sharma}
\affil{Inter-University Center for Astronomy and Astrophysics \\
Pune, Maharashtra 411007\\
India}

\author[0000-0003-3220-7543]{Shabnam Iyyani}
\affil{Inter-University Center for Astronomy and Astrophysics \\
Pune, Maharashtra 411007\\
India}

\author[0000-0003-3352-3142]{Dipankar Bhattacharya}
\affil{Inter-University Center for Astronomy and Astrophysics \\
Pune, Maharashtra 411007\\
India}



\begin{abstract}
The nature of the gamma-ray burst (GRB) central engine still remains an enigma.  Entities widely believed to be capable of powering the extreme jets are magnetars and black holes. The maximum rotational energy that is available in a millisecond magnetar to form a jet is $\sim 10^{52}\, \rm erg$. We identify $8$ long GRBs whose jet opening angle corrected energetics of the prompt emission episode are $> 10^{52}\, \rm erg$ with high confidence level and therefore, their central engines are expected to be black holes. Majority of these GRBs present significant emission in sub-GeV energy range. The X-ray afterglow light curves of these bursts do not show any shallow decay behaviour such as a plateau, however, a few cases exhibit flares and multiple breaks instead of a single power-law decay. For a minimum mass of the black hole ($\sim 2 M_{\odot}$), we find the efficiency of producing a jet from its rotational energy to range between $2\% - 270\%$. Highly energetic jets requiring high efficiencies implies that either the mass of these black holes are much larger or there are, in addition, other sources of energy which power the jet. By considering the Blandford-Znajek mechanism of jet formation, we estimate the masses of these black holes to range between $\sim 2 - 60 \, \rm  M_{\odot}$. Some of the lighter black holes formed in these catastrophic events are likely candidates to lie in the mass gap region ($2 - 5 \, M_{\odot}$).  

\end{abstract}



\section{Introduction} 
\label{sec:intro} 

GRBs are extremely luminous sources with isotropic equivalent energies ranging between $10^{47}-10^{55}$ erg \citep{Ajello_etal_2019}. These observations suggest that the central engine of these bursts should be capable of launching highly energetic jets which significantly exceed the Eddington luminosity. The smallest time variability observed in GRB lightcurves is of a few milliseconds which suggests emission originating from compact sources of radius of the order of $\sim 10^{8}-10^{9}$ cm \citep{MacLachlan_etal_2013}. Broadly, two types of central engines are considered for GRBs:
(i) a hyper-accreting stellar-mass black hole \citep{woosley1993gamma,narayan2001accretion,mckinney2005jet}, and (ii) a rapidly spinning, highly magnetized, neutron star (NS) or `fast magnetar' \citep{Usov1992,Duncan_Thompson1992,Metzger_etal_2011}.

Generally, inferences regarding the plausible central engine of GRBs are made by studying the various features such as plateau, flares and steep decays observed in the X-ray afterglow flux light curves detected by {\it Neil Gehrels Swift} Observatory's X-Ray Telescope (XRT) \citep{Rowlinson_etal_2014,Nathanail_etal_2016,Lei_etal_2017,
Liang_etal_2018,Sarin_etal_2019,Zhao_etal_2020}.
However, such piecemeal study of the afterglow light curves leaves several unanswered questions regarding the mechanism of powering the relativistic jets. Both magnetar, as well as black hole central engine models, can explain most of the features present in the XRT lightcurves, which results in ambiguity and uncertainty regarding the central engine of the GRB. On the other hand, a robust method is to compare the energetics of the GRB with both the central engine models \citep{cenko2011}.

Magnetars are neutron stars with high magnetic fields ($10^{15}\; G$) \citep{Usov1992,Duncan_Thompson1992,Metzger_etal_2011}. To power a GRB, a magnetar must also be spinning rapidly. The spin frequency distribution of accreting millisecond X-ray pulsars are found to show a sharp cutoff at $730\, \rm Hz$ which corresponds to a periodicity, $P_{ns} \sim 1 \, \rm ms$ \citep{fastest_pulsar2006,Chakrabarty2008,Papitto_etal_2014,Patruno_etal_2017}. The mean of the mass distribution of millisecond pulsars is found to be $M_{ns} = 1.48 \pm 0.2 \, M_{\odot}$ \citep{Ozel_etal_2012}. The equation of state of neutron star 
gives a corresponding maximum possible radius ($R_{ns}$) of the neutron star to be $\sim 12  \, \rm km$.    
The rotational energy of the magnetar that powers the GRB jet, normalised to typical observed parameters is estimated as to be,
\begin{equation}
  E_{\rm rot} \simeq \frac{1}{2} \; I \; \Omega^2 \approx 3\times10^{52} erg \left(\frac{M_{ns}}{1.5 \; M_{\odot}}\right) \left(\frac{R_{ns}}{12\;km}\right)^2 \left(\frac{P_{ns}}{1ms}\right)^{-2}
  \label{eq:mag_rot}
\end{equation}
where, I is the moment of inertia of neutron star calculated as $\frac{2}{5} M_{ns} R_{ns}^2$ and $\Omega$ is the rotational speed. More massive NSs with $M_{ns} = 2 M_{\odot}$ can have $P_{ns}$ as small as $0.7$ ms, and $E_{rot}$ may reach $\sim 10^{53}$ erg. Such a short period is not yet observed in any millisecond pulsar. Also, the maximum rotational energy from the magnetar decreases rapidly when the magnetar mass is above the maximum stable mass limit of $\sim 2.1 - 2.4 M_{\odot}$ (supramassive neutron stars; refer Figure 8 in \citealt{Metzger2017}). 
In addition,
\cite{Metzger_etal_2018} (also see \citealt{Beniamini_etal_2017}) has shown that when considering scenarios like the fall-back accretion on to a magnetar, the process of accretion does not significantly alter the maximum rotational energy that is available from a magnetar in comparison to  an isolated (non-accreting) magnetar, that is spinning near the breakup value of $1$ ms. Thus, theoretically, moderately higher values of $E_{rot}$ can be expected from the magnetar, however, 
while considering realistic scenarios like fall-back accretion and
based on the observation of the various parameters like mass and periodicity of the neutron stars, we find the $E_{rot}$ to be roughly around a few times $10^{52}$ erg (consistent with equation \ref{eq:mag_rot}). 

The rotational energy of the magnetar apart from powering the GRB jet, is also lost through magnetospheric winds and gravitational waves. Thus, by considering equi-partition\footnote{In a more realistic scenario, only a small fraction of the rotational energy is expected to be converted into the jet \citep{meszaros2006gamma} and therefore, the burst energy limit of $10^{52}\, \rm erg$ presents an elevated upper limit for the jet produced by a magnetar.}, we find it very reasonable to consider that  
the maximum extractable rotational energy of magnetar that can be channelised into powering a relativistic jet is $E_{\rm rot,jet} = 1 \times 10^{52}\, \rm erg$.
Any GRB with a total burst energy exceeding this energy budget can be, thus, considered to not possess a magnetar but instead a black hole as its central engine. 

In this paper, we use the burst energetics of the prompt gamma-ray emission of the GRBs to identify the bursts with black hole as their central engines. The bursts' energy calculations done throughout the paper uses the standard $\Lambda_{\rm CDM}$ cosmology, with cosmological parameters, $H_{0} = 67.4 \pm 0.5$ km/s/Mpc, $\Omega_{vac} = 0.685$ and $ \Omega_{m} = 0.315$  (Planck collaboration 2018, \citealt{2018_planck}).

\section{Sample Selection}
\begin{figure}
\centering
	\includegraphics[scale=0.51]{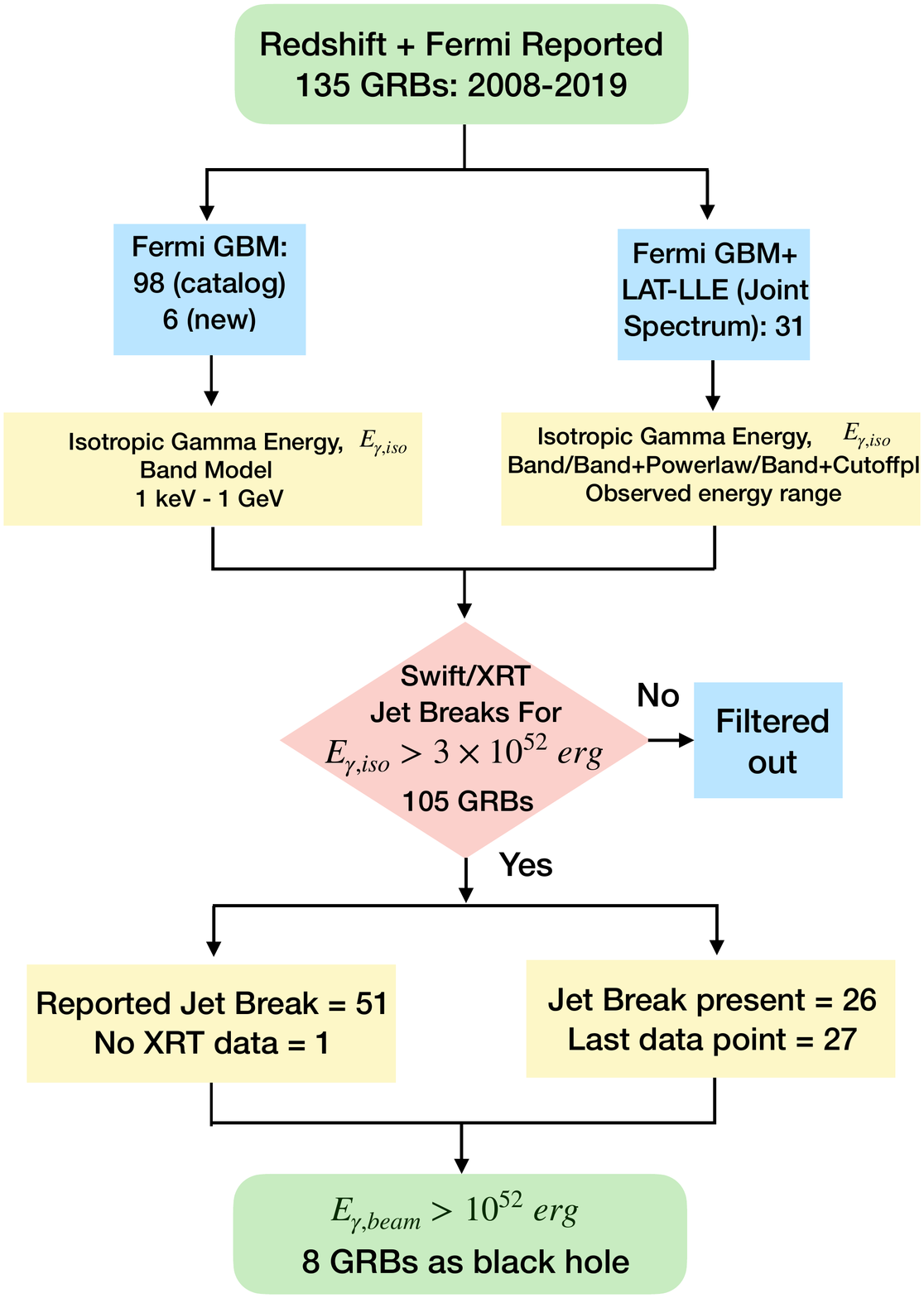}
    \caption{Flowchart of sample selection process}
    \label{fig:flowchart}
\end{figure}

GRBs with known redshifts detected by {\it Fermi} gamma-ray space telescope\footnote{https://heasarc.gsfc.nasa.gov/W3Browse/fermi/fermigbrst.html} are extracted from the catalogue presented in the online GRB table of Jochen Greiner, which is available at {\it http://www.mpe.mpg.de/~jcg/grbgen.html}. 
{\it Fermi} observations provide a spectral coverage spanning over several decades of energy between a few keV to several GeV. 
This allows us to model the spectrum of the prompt emission better and estimate the bolometric energy flux of the burst.
Our sample consists of $135$ GRBs with redshift information and detected by {\it Fermi} during the years 2008 - 2019. 
The various steps undertaken to identify the GRBs with black hole central engines are described in the section below and are also summarised in the flowchart presented in Figure~\ref{fig:flowchart}.

\section{Methodology of identification}

\subsection{Isotropic prompt gamma-ray emission, $E_{\gamma, iso}$}
Assuming an isotropic emission, the total gamma-ray energy released during the prompt phase of the burst is termed as $E_{\gamma, iso}$. The isotropic equivalent energy is found as \citep{bloom2001prompt},
\begin{equation}
    E_{\gamma,iso} = \left(\frac{4\;\pi\;D_{L}^{2}}{1+z}\right) \; F_{bol}
    \label{eq:energy_bol}
\end{equation}
where, $F_{bol}$ is the bolometric fluence and $D_{L}$ is the luminosity distance of the burst at a redshift, $z$.

In total, there are $104$ GBM~only detected and $31$ LAT + GBM detected GRBs with known redshifts. For this work, we conducted the time-integrated spectral analyses of all the $31$ LAT detected GRBs and $6$ GBM only detected GRBs whose spectral parameters are not updated in the {\it Fermi} catalogue. The spectral analyses of the GBM~only cases are done using the \texttt{Band} model, whereas, the joint time-integrated spectral analyses using the GBM and LAT data are performed using various models such as \texttt{Band}, \texttt{Band + Power-law} and \texttt{Band + Cutoff-powerlaw}. The best fit model is then used for the fluence estimation. In {\it Fermi}/GBM catalogue, the fluence is reported for $10-1000$ keV energy range which gives an underestimation of the total energy released during the burst. Therefore, the bolometric fluences of all the bursts are estimated within the energy limits\footnote{In LAT detected cases, wherever the highest energetic photon detected exceeded 1 GeV, the fluence was estimated in the energy range extending up till that highest observed energy value.} of $1$ keV and $1$ GeV. Note that for $98$ GBM only detected GRBs, the bolometric fluences are estimated using the spectral parameters for the \texttt{Band} function fits and the $T_{90}$ given in the {\it Fermi} catalogue. 

The isotropic equivalent burst energies of the $135$ GRBs thus obtained are then compared with the maximum possible rotational energy limit of a magnetar ($3 \times 10^{52} \, \rm erg$) and a total of $105$ hyper-energetic GRBs are found to exceed this energy budget (Figure~\ref{fig:Energy_z}a). We note that no short GRB made into this list of hyper-energetic GRBs.

\subsection{Beam corrected prompt emission, $E_{\gamma, beam}$}
GRB outflows are collimated relativistic jets which means the exact burst energy is the amount of energy that is ejected into the solid angle forming the jet. This is referred to as the beam corrected prompt emission, $E_{\gamma, beam}$.
This is estimated by multiplying the isotropic burst energy, $E_{\gamma, iso}$, with the beaming correction factor \citep{frail2001beaming} given as,
\begin{equation}
    f_b = 1-cos\;\theta_j
    \label{eq:corr}
\end{equation}
where $\theta_j$ is the opening angle of the jet.
In this work, we consider only the uniform (top-hat) 
jet scenario which is the typical model considered in 
literature (see Appendix \ref{caveats} for discussion on structured jets).

The jet opening angle values for bursts that are already reported in the literature until June 2020 are used as it is ($51$ cases\footnote{We have used the beaming angles reported in literature, which were estimated by modelling the optical, radio or late time X-ray afterglow observations by considering a uniform jet with on-axis and off-axis (particularly, \citealt{zhang2015}) viewing geometry. In the case of GRBs, where multiple values of $\theta_j$ are reported, we have used either the well-constrained value or the lower estimate as that would provide a more stringent constraint on the energy.}) and in the remaining cases, the jet opening angle is estimated by using the time of jet breaks observed in {\it Neil Gehrels Swift} Observatory/XRT afterglow observations ($26$ cases). In cases where the jet break is not observed ($27$ cases), the last data point in the XRT observation is used to estimate the lower limit of the possible jet opening angle of that GRB \citep{sari1999_jets,frail2001beaming}.  In Appendix \ref{caveats}, we have discussed the different caveats that are involved in the $E_{\gamma,beam}$ estimation.

\subsubsection{Jet opening angle calculation}
The online XRT repository\footnote{$https://www.swift.ac.uk/xrt\_curves/$} is used for extracting the energy flux light curves in $0.3-10$ keV energy range and the time of the jet break. The XRT products are created using automatic analysis described in \cite{evans2007online, evans2009methods}. Using the above information, the jet opening angle is estimated under the assumption of standard afterglow model, on-axis viewing geometry and a uniform jet, by the following expression \citep{sari1999_jets, wang2015_break},

\small
\begin{equation}
    \theta_{j} \approx 0.057  \left(\frac{t_j}{1\;day}\right)^{3/8}  \left(\frac{1+z}{2}\right)^{-3/8}  \left(\frac{E_{\gamma,iso}}{10^{53} \; erg}\right)^{-1/8}  \left(\frac{\epsilon}{0.2}\right)^{1/8}  \left(\frac{n_{p}}{0.1\;cm^{-3}}\right)^{1/8}
    \label{eq:jet_break}
\end{equation}
\normalsize
where $t_j$ is the time of jet break in days, $\epsilon$ is a measure of how efficiently the total energy of the burst is converted into radiation and $n_{p}$ is ambient medium density.
The estimate of $\theta_{j}$ is weakly dependent on $\epsilon$ and $n_{p}$ which are largely unknown. Following the methodology in \citet{goldstein2016estimating}, we assume a broad uniform distribution for $\epsilon$ from $5\%$ to $95\%$, considering the earlier reported range of radiation efficiencies \citep{cenko2011,Racusin_etal_2011}. Based on the limited number of estimates made for $n_p$ previously, a log-normal distribution with mean $log_{10}(0.1)$ and standard deviation $1$ is assumed. In this work, we have also considered a uniform distribution for $t_j$ within its uncertainty limits obtained from the observations.
The probability distribution of $\theta_{j}$ is, thus, built by evaluating $\theta_j$ (Eq~\ref{eq:jet_break}) for each Monte Carlo sampled set of values of $t_j$, $\epsilon$ and $n_{p}$ from their respective probability distributions. By fitting a Gaussian distribution function to the obtained distribution, 
we obtain the mean value of $\theta_{j}$ and the standard deviation as its uncertainty. The distribution of $\theta_j$ obtained for the GRBs where the jet break is observed or reported earlier is shown in Figure~\ref{fig:Energy_z}b. 

In case of GRBs, where the jet break is not observed in XRT energy flux lightcurve, using the time of the last data point in the lightcurve, we estimate the distribution of lower limit of $\theta_j$ for the GRB using the above mentioned assumptions and Monte Carlo method.
The value obtained after subtracting the standard deviation from the mean of the distribution is considered as $\theta_{j, min}$ for the GRB. 

\subsection{Result}
\begin{figure}
\centering
	\includegraphics[scale=0.5]{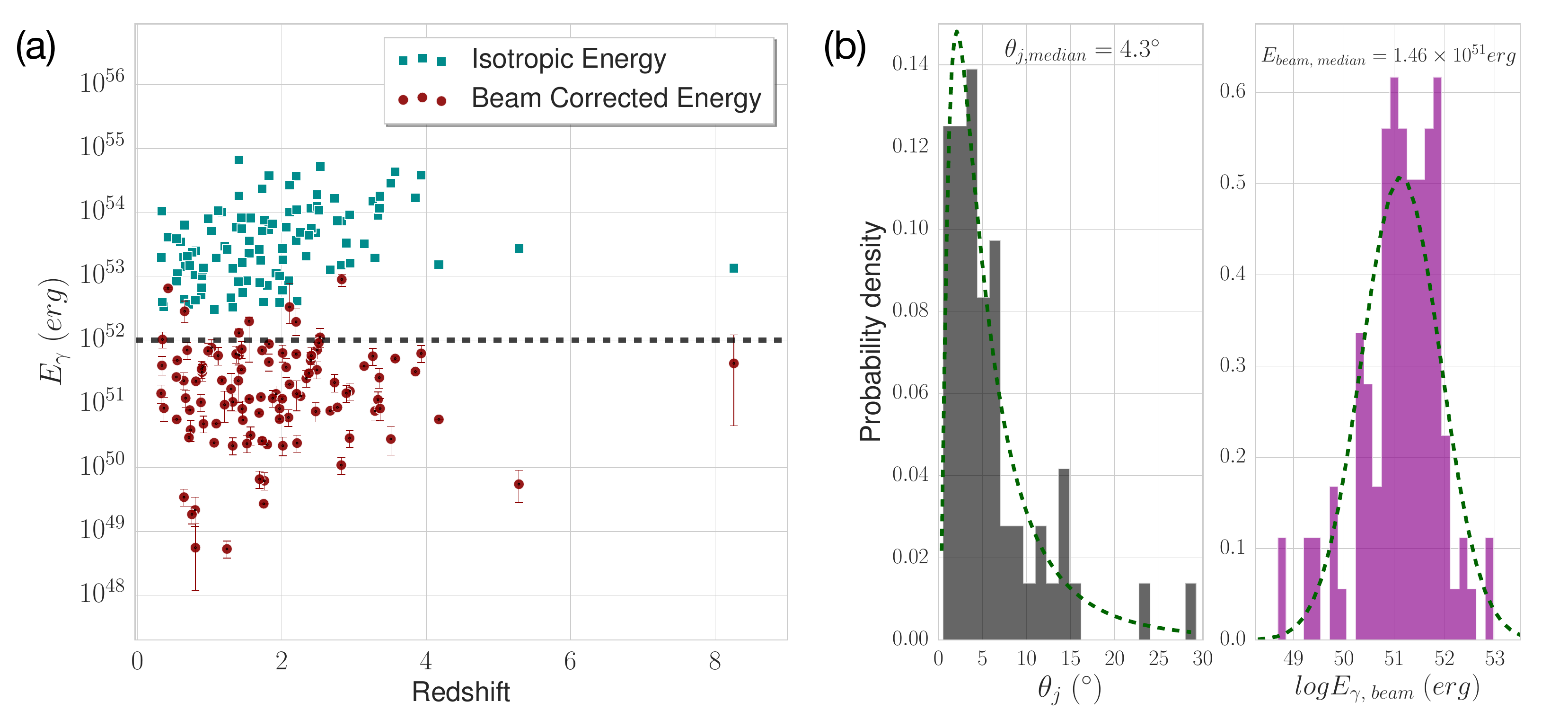}
    \caption{(a) The $E_{iso}$ and $E_{\gamma,beam}$ 
    of the prompt emission of the $105$ hyper-energetic GRBs are shown in green squares and red circles respectively. The horizontal grey dashed line marks $10^{52}\, \rm erg$. (b) The distributions of $\theta_j$ 
    and 
    $E_{\gamma,beam}$ of the $105$ hyper-energetic GRBs are shown in the left and right panels of the plot respectively. The log-normal fits to the respective distributions are shown in dashed green lines. The means of $\theta_j$ and $E_{\gamma,beam}$ distributions are $2.1^{\circ} \pm 0.9^{\circ}$ and $(1.9 \pm 4.2) \times 10^{51}\, \rm erg$ respectively.}
    \label{fig:Energy_z}
\end{figure}

The isotropic equivalent energies observed in prompt gamma-ray emission ($E_{\gamma, iso}$) for the $105$ GRBs, which exceed the magnetar energy budget ($10^{52}$\, \rm erg) are shown in Figure~\ref{fig:Energy_z}a. After beam correction, we observe a wide distribution of beam corrected energies, $E_{\gamma, beam}$ in the prompt phase (Figure~\ref{fig:Energy_z}a). The distribution of the  $E_{\gamma, beam}$ of these GRBs is shown in Figure~\ref{fig:Energy_z}b. 

The total burst energy is the sum of $E_{\gamma, beam}$ and the remaining kinetic energy of the jet estimated from the afterglow emissions. By using just the prompt gamma-ray energetics, we find $8$ long GRBs whose $E_{\gamma, beam}$ exceed or are nearly equivalent to the limit of the maximum possible rotational energy of the magnetar that can be converted into a jet ($1 \times 10^{52}\, \rm erg$). Following \cite{Racusin_etal_2011,cenko2011}, it is reasonable  to consider that these bright {\it Fermi} detected GRBs possess high radiation efficiencies ($\epsilon >0.2$), which in turn suggests that the total burst energies in these GRBs can be $E_{burst} \ge E_{\gamma, beam}$.
This assures that the total burst energies of these GRBs are even greater than this limit ($> 10^{52}\, \rm erg$) and thereby confirm that the central engine or the remnant of the core-collapse of the massive progenitor star of these GRBs are black holes. These GRBs along with their observational features are listed in Table~\ref{bh_sample}. 

\begin{table}
\let\nobreakspace\relax
\begin{small}
	\caption{Properties of hyper-energetic GRBs}
	\label{bh_sample}
	\begin{center}
	\hspace*{-2.5cm}
	\begin{tabular}{p{0.98cm}cccccccccc}
        \hline \hline
		GRB name    & $T_{90}$ (\textit{Fermi})  & z      & Fluence        & $E_{\gamma, iso}$             & $\theta_j$\tablenotemark{a} & $\epsilon$\tablenotemark{a}                
		& $E_{\gamma,beam}$ & Confidence\tablenotemark{b}       &  $M_{*}/M_\odot$  & \textit{Swift}/\\
		            & (s)                       &         & $10^{-4} \; erg/cm^2$           & $10^{52} \; erg$              & $^{\circ}$           &     & $10^{52} \;erg$ & level & & XRT Feature\\
		\hline
		190114C     & 116.354           & 0.425  & $8.5_{-0.3}^{+0.3}$           & $41.2_{-1.3}^{+1.4}$       & $>32.5$                   & $0.18$ 
		& $> 6.5_{-0.2}^{+0.2}$ & $>99.99\%$ & $40 - 60$   & 3 Breaks\\ 
		\noalign{\vskip 2.1mm}
		
		180720B     & 48.897            & 0.654  & $5.4_{-0.4}^{+0.5}$           & $63.3_{-5.2}^{+5.7}$       & $> 17.2_{-2.6}^{+2.6}$      
		& -- 
		& $> 2.8_{-0.9}^{+1.2}$ & $99.6\%$&  $5 - 7$ & Flare, 3 Breaks\\ 
		\noalign{\vskip 2.1mm}
		
		170214A\tablenotemark{c}     & 122.882           & 2.53   & $3.5_{-0.1}^{+0.2}$           & $525.7_{-21.0}^{+23.0}$    & $> 3.7_{-0.6}^{+0.6}$        
		& -- 
		& $> 1.1_{-0.3}^{+0.4}$ & $61.5\%$ & $2.14 - 3$  & Straight line\\ 
		\noalign{\vskip 2.1mm}
		
		160625B\tablenotemark{c,d}     & 453.385           & 1.406  & $12.4_{-0.4}^{+0.4}$          & $657.8_{-20.6}^{+22.2}$    & $3.6_{-0.2}^{+0.2}$  & $0.98$     
		& $1.3_{-0.2}^{+0.2}$ & $98.3\%$ & $2.14 - 2.22$  & 1 Break\\ 
		\noalign{\vskip 2.1mm}
		
		

		120624B     & 271.364           & 2.197  & $3.1_{-0.5}^{+0.6}$          & $371.1_{-57.4}^{+74.6}$    & $> 5.8_{-0.9}^{+0.9}$       
		& --  
		& $> 1.9_{-0.7}^{+1.1}$ & $94.6\%$ & $3 - 5$  & A few points\\ 
		\noalign{\vskip 2.1mm}
		
		110731A     & 7.485             & 2.83   & $0.4_{-0.1}^{+0.1}$          & $72.9_{-13.2}^{+14.5}$    & $28.9_{-0.7}^{+0.0}$  & 0.86    
		& $8.9_{-2.0}^{+1.8}$ & $99.9\%$ & $12 - 17$ & Flare, 3 Breaks\\ 
		\noalign{\vskip 2.1mm}
		
		090926A\tablenotemark{e}     & 13.76             & 2.106  & $2.4_{-0.03}^{+0.03}$        & $267.4_{-26.2}^{+33.8}$   & $9_{-2}^{+4}$          & $0.98$    
		& $3.3_{-1.5}^{+4.4}$ & $90.8\%$ & $4 - 6$ & Straight line\\ 
 		\noalign{\vskip 2.1mm}
 		
 		090102       & 26.624            & 1.547  & $0.4_{-0.1}^{+0.1}$          & $22.8_{-1.7}^{+1.6}$      & $23.9_{-12.1}^{+1.1}$  & 0.25     
 		& $2.0_{-1.5}^{+0.3}$ & $85\%$ & $9 - 13$ & 1 Break\\ 
		\noalign{\vskip 2.1mm}
		
        \hline \\
    \end{tabular}
    \vspace{-0.5cm}
\tablenotetext{}{All the errors reported above are $68\%$ confidence intervals of the estimated parameters.}
\tablenotetext{a}{The references are provided for the $\theta_j$ values; and the kinetic energy estimates that are used to evaluate the $\epsilon$ values that are adopted in this work. 
GRB 190114C - \citealt{misra2019}; GRB 180720B - This work; GRB 170214A - This work; GRB 160625B - \citealt{alexander2017} ( also see \citealt{Cunningham_etal_2020} where $\theta_j$ estimation ranges from $1.26$ to $3.90$ deg and can lower the beam corrected total energy of the burst.)}; GRB 120624B - This work; GRB 110731A - \citealt{zhang2015}; GRB 090926A - \citealt{cenko2011}, GRB 090102 - \citealt{zhang2015}.
\tablenotetext{b}{The confidence levels of the reported values/ lower limits of the $E_{\gamma,beam}$ to lie above the considered energy budget limit of $10^{52}\, \rm erg$ are listed. The probability distribution of $E_{\gamma, beam}$ of each GRB is generated by randomly drawing the parameters from their respective Gaussian distributions of $E_{\gamma, iso}$ and $\theta_j$ with value and its error as mean and standard deviation for a million runs. The obtained probability distribution is used to estimate the confidence interval such that the probability of $E_{\gamma, beam}>10^{52} erg$.}

\tablenotetext{c}{In these GRBs, the possible lower limit of black hole mass is considered to be $2.14 \, M_{\odot}$ as the estimated lower limits are less than the minimum possible mass of stellar mass black holes.}

\tablenotetext{e}{GRB 090926A was the only hyper-energetic {\it Fermi} GRB that was identified in \cite{cenko2011} to pose a severe challenge to the magnetar central engine model.}

\end{center}
\end{small}
\end{table}

%


\section{Discussion}
Below, we discuss the various properties of the prompt, afterglow emissions, and the black hole central engine of the 8 long GRBs. 

\subsection{Sub-GeV loud}
All, but one (GRB 090102), of the $8$ long GRBs show significant emission in the LAT energy range of $30 \, \rm MeV$ to several GeV (see the light curves in the Figure \ref{lcs_1} and \ref{lcs_2} in the Appendix). 
We note that there is no redshift preference for these LAT detected GRBs (see Figure~\ref{fig:Energy_z}, also refer \citealt{Ajello_etal_2019}), which implies that long GRBs with such high energy emissions are produced in different epochs of the universe. This affirms the positive correlation between the strong LAT emission and the high burst energetics which may further imply that the central engines of such highly energetic bursts are most likely black holes.    

\subsection{X-ray afterglow lightcurves}
{\it Neil Gehrels Swift} Observatory/XRT can be slewed to the target within tens of seconds and hence provides observations of the early afterglow phase of GRBs in X-rays.
The XRT observations have revealed features like flares, plateau, steep decay etc in the flux light curves, which are related to the continued activity of the central engine well beyond the timescale of the prompt emission \citep{Yamazaki_etal_2020}. In several studies, these features are explained within the framework of both black hole \citep{Pawan_etal_2008,Nathanail_etal_2016,Lei_etal_2017} and magnetar \citep{Barniol_Kumar2009,Rowlinson_etal_2014,Liang_etal_2018,Sarin_etal_2019,Zhao_etal_2020} central engine models. Generally, the observance of plateau and the steep decay thereafter have been interpreted as potential signatures of magnetar where the plateau is produced by the energy injection from the magnetospheric wind, whereas the post-plateau steep decay signifies the collapse of the magnetar to a black hole \citep{Rowlinson_etal_2014,Bernardini_2015_magnetar,Chen_etal_2017_magnetar_to_BH}.    

\begin{figure}
\centering
	\includegraphics[scale=0.54]{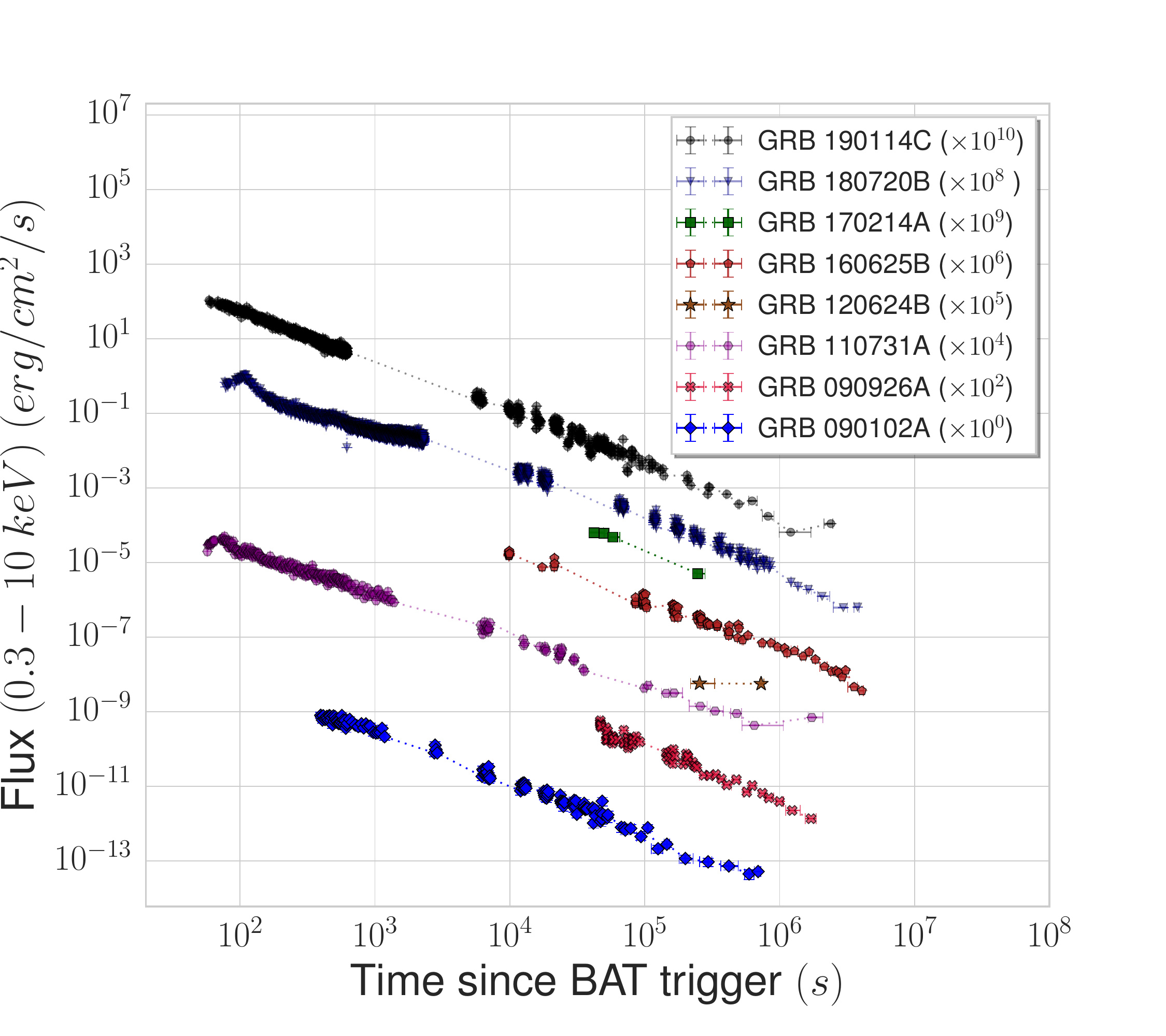}
    \caption{The X-ray afterglow lightcurves of the 8 long GRBs with black hole central engines, observed by the XRT instrument are shown. }
    \label{fig:xrt_lc}
\end{figure}

Since the X-ray afterglow light curves are considered to shed some light on the central engine, we have extracted the XRT flux lightcurves\footnote{XRT online repository, $https://www.swift.ac.uk/xrt\_curves/$} of these black holes candidates and shown them in Figure~\ref{fig:xrt_lc}. The observational features in XRT light curves for the black hole cases are reported in Table~\ref{bh_sample}. 
We note the following key points in these X-ray light curves: (i) neither shallow decay feature like plateau nor steep decays are observed; (ii) flares are observed in two cases as early as less than a few hundred seconds; (iii) apart from the jet break, multiple other breaks are observed in the flux light curves.

\subsection{Properties of black hole central engine}
With the black hole as the central engine, the powering of the
jet can happen via two main mechanisms such as neutrino 
annihilation in a neutrino-dominated accretion flow 
\citep{Ruffert_etal_1997, Chen_Beloborodov_2007} or by 
extracting the rotational energy of the Kerr black hole 
\citep{Lee_etal_2000, alexander2017}. However, the first process 
has not been found to produce ultra-relativistic GRB jets \citep{Leng_Giannios_2014} successfully.      
We, therefore, consider that the observed prompt emission is 
dominantly powered by the rotational energy ($E_{rot}$) of the 
black hole central engine, such that 
\begin{equation}
    \eta \, E_{rot} = E_{\gamma,beam}
    \label{eq:Erot_beam}
\end{equation}
where $\eta$ represents the net efficiency of converting the 
rotational energy of the black hole into the observed gamma-ray 
emission of the GRB and 
\begin{equation}
    E_{rot} =  f(a_{*}) \; \frac{M_{*}}{M_\odot} \; c^{2} = 1.8 \times 10^{54} \; f_{rot}(a_{*}) \; \frac{M_{*}}{M_\odot} erg
    \label{eq:spin_energy}
\end{equation}
where $M_{*}$ is the mass of the black hole and
\begin{equation}
    f(a_{*}) =  1 - \sqrt{(1+\sqrt{1-a_{*}^2})/2}
    \label{eq:spin}
\end{equation}
where $a_{*}$ = $Jc /GM_{*}^2 $ is the dimensionless black hole 
spin parameter, $J$ is the angular momentum of the black hole.

\subsubsection{Jet powering efficiency, $\kappa$}
The efficiency $\eta$ is dominated by two main factors: (i) the 
fraction ($\kappa$) of the $E_{rot}$ that is channelled into 
powering the relativistic jet, which depends on the 
mechanism of how the rotational energy is extracted, and (ii) 
the fraction of the jet power ($\epsilon$) that is eventually 
radiated away in gamma-rays only. In the five cases where the kinetic energy estimates of the bursts that are evaluated using the multi-wavelength data of the afterglow observations are available in literature, we have used those values to estimate the respective radiation efficiency ($\epsilon$) of the bursts. In three GRBs where the kinetic energy estimates of the bursts are not available in literature, we adopted the average ($\epsilon=0.65$) of the $\epsilon$ values found for the other five GRBs (Table \ref{bh_sample}). We find this average value of $\epsilon$ to be consistent with the previous studies of
the estimates of radiation efficiencies of hyper-energetic GRBs detected by {\it Fermi}
\citep{Racusin_etal_2011,cenko2011}. 
The hyper-accreting black holes formed during the GRB events are
considered to be initially moderately spinning with $a_* \sim 
0.5$ which later spins up close to maximal spin of $a_* \sim 
0.9$ \citep{Narayan_etal_1992,MacFadyen_Woosley_1999,Shapiro_Shibata_2002,Shibata_Shapiro_2002}. In core-collapse of massive 
stars, stellar-mass black holes are formed when the remnant core
exceeds the maximum possible mass of a stable neutron star that 
can be formed. The maximum mass of a neutron star observed till 
date is $2.14 \, M_{\odot}$ \citep{Cromartie_etal_2020_maxNSmass}.  

\begin{figure}
\centering
    \includegraphics[scale=0.5]{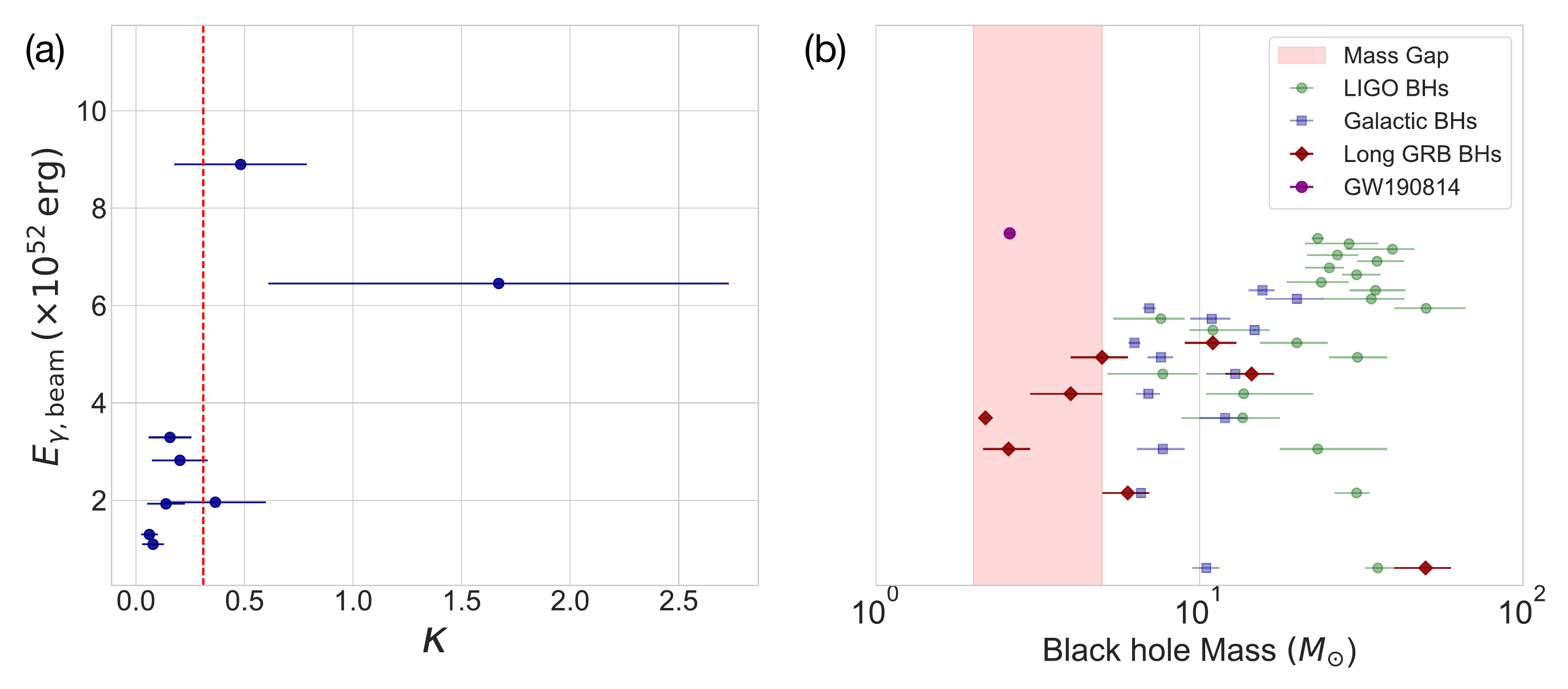}
    \caption{(a) The GRB prompt emission energy, $E_{\gamma,beam}$ versus the jet powering efficiency, $\kappa$ is shown. The possible parameter space of $\kappa$ is represented by the limits of the error-bar, the mean of these limits are marked by the solid black circles. The maximum efficiency of BZ mechanism is marked by the red dashed line. (b) The estimated parameter space of the mass of the black hole central engines of the 8 long GRBs listed in the Table~\ref{bh_sample} are represented by the limits of the error-bars, the mean of these limits are marked by the dark red diamonds. For comparison, masses of the black holes observed by LIGO in binary black hole mergers \citep{Abbott_etal_2019}, and the masses of the galactic black holes estimated in X-ray binaries \citep{Wiktorowicz_etal_2014} are shown in green circles and blue squares respectively. The secondary merger component of unknown nature detected by LIGO in GW190814 is shown in purple circle \citep{Abbott2020}.}
    \label{fig:eff_kappa}
\end{figure}

Inserting the above reasonable values for $\epsilon$ and $a_*$ for the minimum possible mass of the black hole, $2.14 \, M_{\odot}$, in the equations \ref{eq:Erot_beam} and 
\ref{eq:spin_energy}, we estimate the parameter space of the jet production efficiency, $\kappa$. The obtained results are shown 
in Figure \ref{fig:eff_kappa}a. The high burst energetics require that a large amount of rotational energy is extracted from the black hole. This is reflected in 
the positive correlation obtained between $\kappa$ and the burst energies of the GRBs.
For some of the brightest GRBs with 
$E_{\gamma,beam} \ge 2 \times 10^{52}\, erg$, we find the upper limits of $\kappa$ to range between $30\%$ and $270\%$. 
The maximum fraction of the rotational energy of a black hole that
can be extracted by a Blandford-Znajek mechanism is $0.31$ 
\citep{Lee_etal_2000}. Therefore, such high values of $\kappa$ implies either of the two possibilities: (i) The masses of the black hole central engines are much larger than $2 \, M_{\odot}$. In other words, for smaller values of $\kappa$ in these cases require that the rest mass energy of the black holes are much larger; or (ii) there are in addition other sources of energy powering the jet.  

\subsubsection{Black hole mass estimate}
Blandford-Znajek (BZ) process has been widely discussed in the 
literature as the potential mechanism to extract the rotational 
energy of a Kerr black hole \citep{Blandford_Znajek_1977, Lee_etal_2000,mckinney2005jet}. In such a scenario, the jet 
powering efficiency, $\kappa$, can be further understood as the 
product of two main factors: (i) The fraction of the rotational energy of the black hole that can be extracted by BZ process. 
Since these bursts are extremely energetic, it is reasonable to assume that the BZ mechanism works at its maximum efficiency, in
other words, it can extract nearly $31\%$ of the rotational
energy of the black hole \citep{Lee_etal_2000}. 
(ii) The fraction of the extracted BZ power that gets channelled into the formation of the GRB jet. Numerical simulation studies show that for a black hole spin of $a_* =[0.5,0.9]$, the efficiency of converting the extracted BZ power to a jet is found to be $[7\%-47\%]$ \citep{mckinney2005jet}.  

Taking into account these different efficiency factors of
$\kappa$ and $\epsilon$, we find $\eta$ to range between $0.4\% - 14\%$ and 
we estimate the mass of the black hole central engine in these $8$ long GRBs using equations \ref{eq:Erot_beam} and \ref{eq:spin_energy}. We find 
the masses of the black holes to range between $2 - 60 \, 
M_{\odot}$. The values are listed in Table~\ref{bh_sample}
and are also plotted in Figure~\ref{fig:eff_kappa}b. 

The observational measurements of the masses of the compact
remnants, post the core-collapse of massive stars and the merger of
compact objects like binary neutron stars or neutron star -
black hole, have shown a 'gap' between the heaviest neutron 
stars and the lightest black holes. This is generally referred 
to as the `mass gap' region which lies between $2 - 5 \, 
M_{\odot}$ \citep{Ozel_etal_2012}. The recent gravitational wave event, GW190814, detected by LIGO signifies the merger of two objects of masses $22.2 -  24.3\, M_{\odot}$ and $2.50 - 2.67\, 
M_{\odot}$ \citep{Abbott2020}. It is, however, uncertain whether the lighter object is a massive neutron star or the lightest black hole. 
We find that the lighter 
black holes estimated in this study have possible masses close 
to the upper limit of the neutron star mass. Thus, we find that 
some of the black holes formed in these catastrophic events of 
GRBs can be the likely candidates to lie in the mass gap region.

\section{Summary}
Despite several decades of extensive studies and observations of gamma-ray bursts, many aspects of the event still remain largely a mystery. One of these is regarding the central engine powering the ultra-relativistic GRB jets whose luminosities exceed the Eddington luminosity by several orders of magnitude. Broadly, the possible central engine is classified into either a magnetar or a black hole. Much work have been done to investigate these possibilities by studying various features such as plateau and its post steep decay, flares present in the X-ray afterglow light curves etc. However, these studies have remained mostly inconclusive and ambiguous, with both magnetar and black hole models being able to explain the observed features. 

One robust way to identify GRBs with black hole central engine is by looking at the energetics of the GRB event. The maximum possible rotational energy of the magnetar that can be converted into a relativistic jet is $\sim 10^{52} \, \rm erg$. In this work, we use this constraint to identify the GRBs whose beam corrected prompt emission energetics exceed this energy budget. Eight long GRBs are found to possess burst energies greater than $10^{52}\, \rm erg$ and thereby central engines that are most likely black holes. We note that these GRBs are extremely bright with significant emission in the sub-GeV energy range. The X-ray afterglow light curves of these GRBs do not show any `plateau' and steep decay like features. Popularly, such features are associated with the activity of a magnetar central engine. So, the non-observance of these features further asserts that the central engines of these GRBs are black holes. Considering that the jet is dominantly powered by the rotational energy of the black hole which is extracted by the Blandford-Znajek mechanism, we estimate the masses of the black holes to range between $2-60 \, M_{\odot}$. We find that the lighter black holes formed in these catastrophic events could be candidates to lie in the mass gap between the heaviest known neutron star and the lightest known black hole. 

\acknowledgments
We would like to thank Prof. Pawan Kumar and Dr. Vikas Chand for the suggestions. This research has made use of {\it Fermi} data obtained through High Energy Astrophysics Science Archive Research Center Online Service, provided by the NASA/Goddard Space Flight Center. This publication uses data from the {\it Swift} mission, archived at the Swift Data Center (SDC) at the Goddard Space Flight Center (GSFC). This work utilized various software such as HEASARC, \textsc{Xspec}, Python, astropy, corner, numpy, scipy, matplotlib, FTOOLS etc.

%






\appendix

\section{Light curves}
\begin{figure}
    \centering
    \includegraphics[width=.49\linewidth]{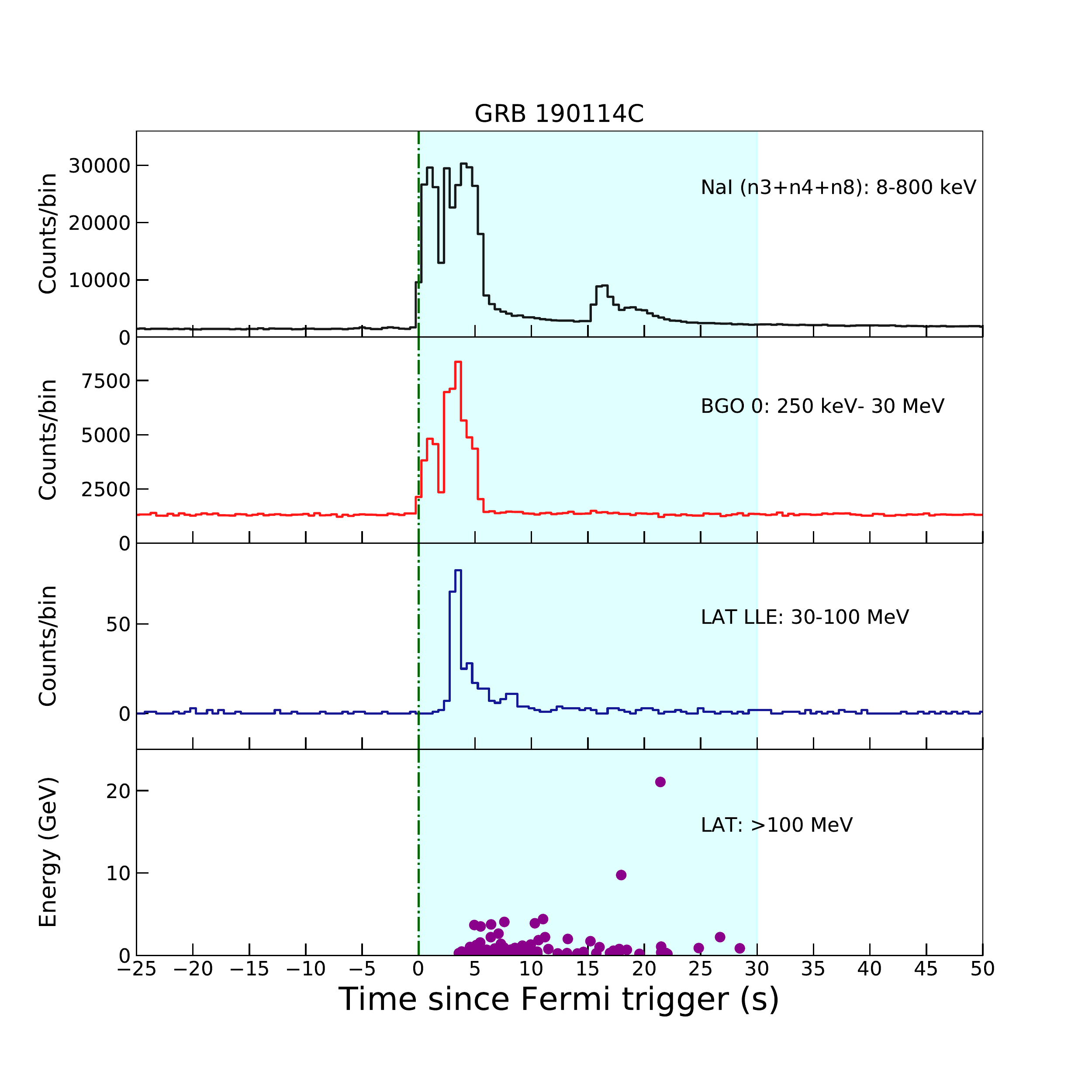}
    \includegraphics[width=.49\linewidth]{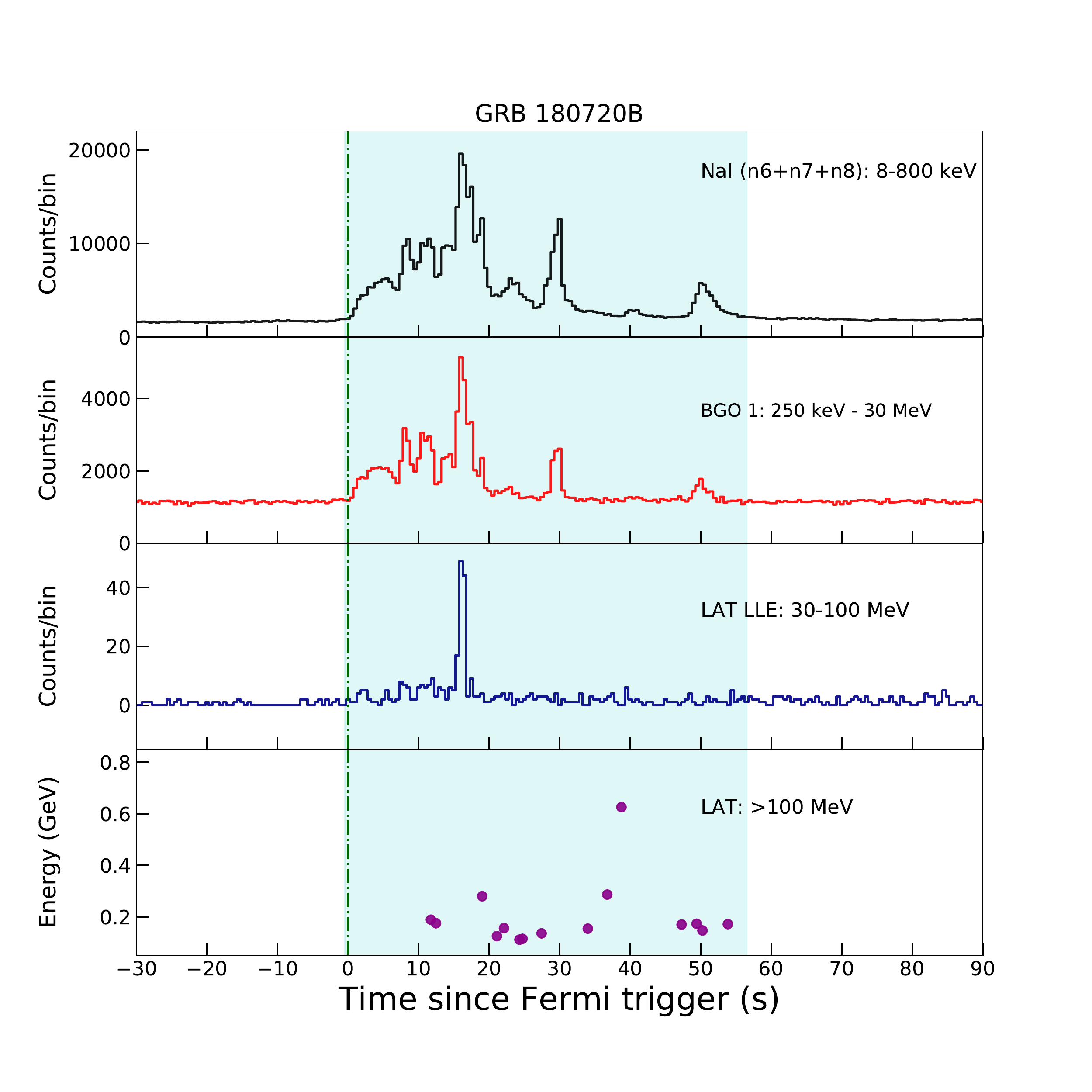}
    \includegraphics[width=.49\linewidth]{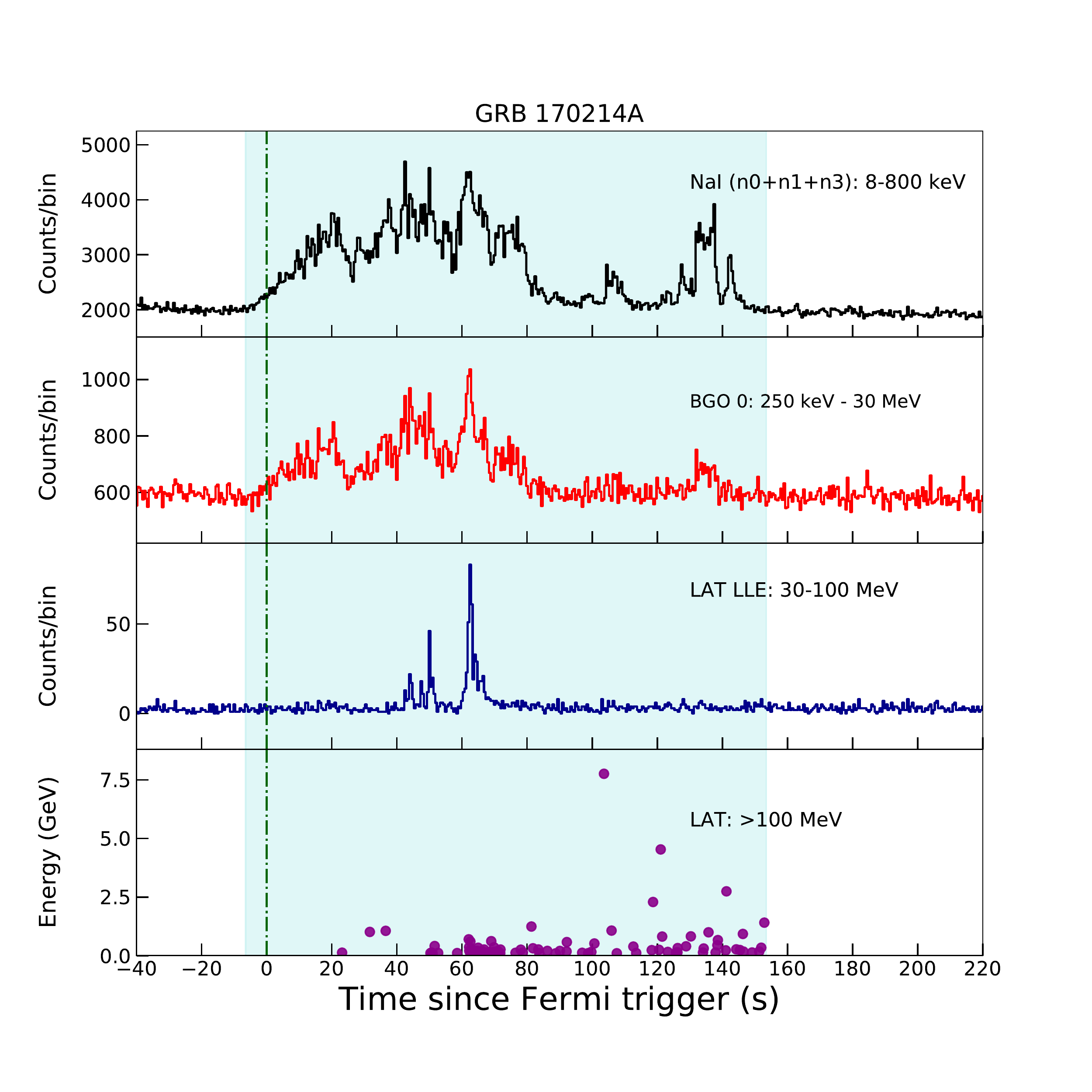}
    \includegraphics[width=.49\linewidth]{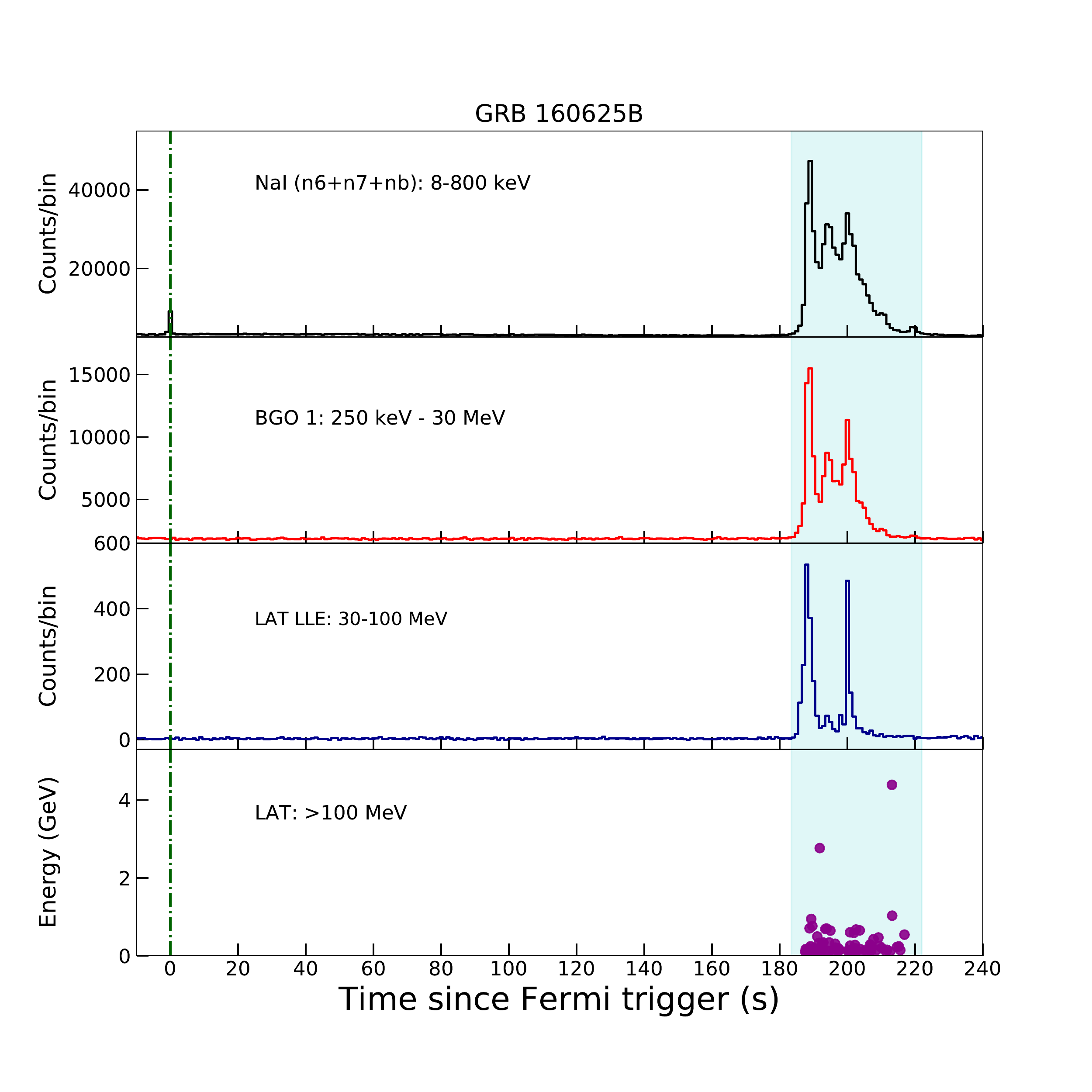}
   \caption{Light curves of the 8 GRBs with the black hole as a central engine: NaI, BGO and LLE light curves are presented in black, red and blue color, respectively from top to bottom. In the bottom panel, of each GRB light curve, LAT photons are shown in magenta color with energy (in GeV) information on the y-axis. The cyan shaded region marks the time interval used for time integrated spectral analysis, and green vertical line represents the trigger time of the GRB.}
    \label{lcs_1}
\end{figure}

\begin{figure}
    \centering
    \includegraphics[width=.49\linewidth]{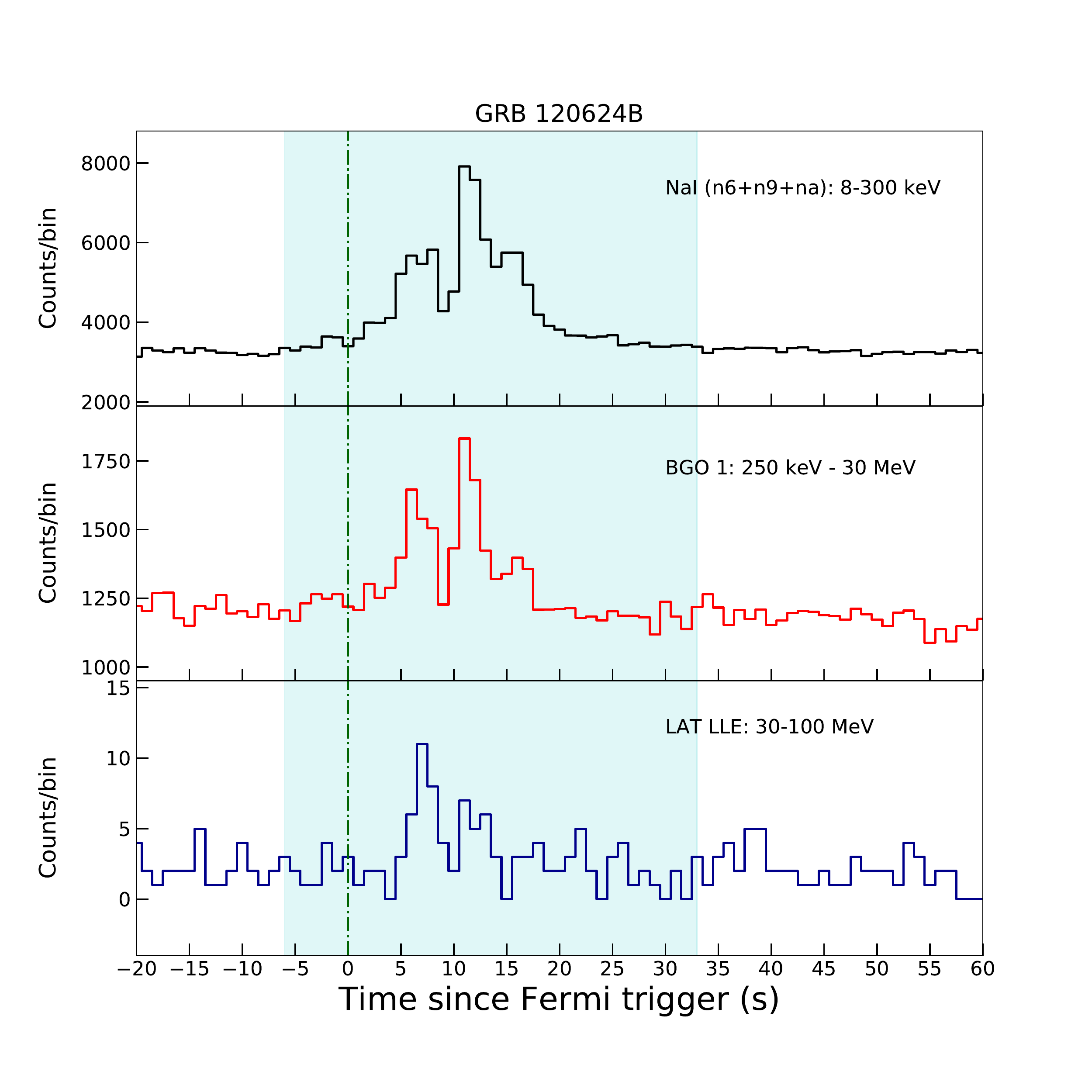}
    \includegraphics[width=.49\linewidth]{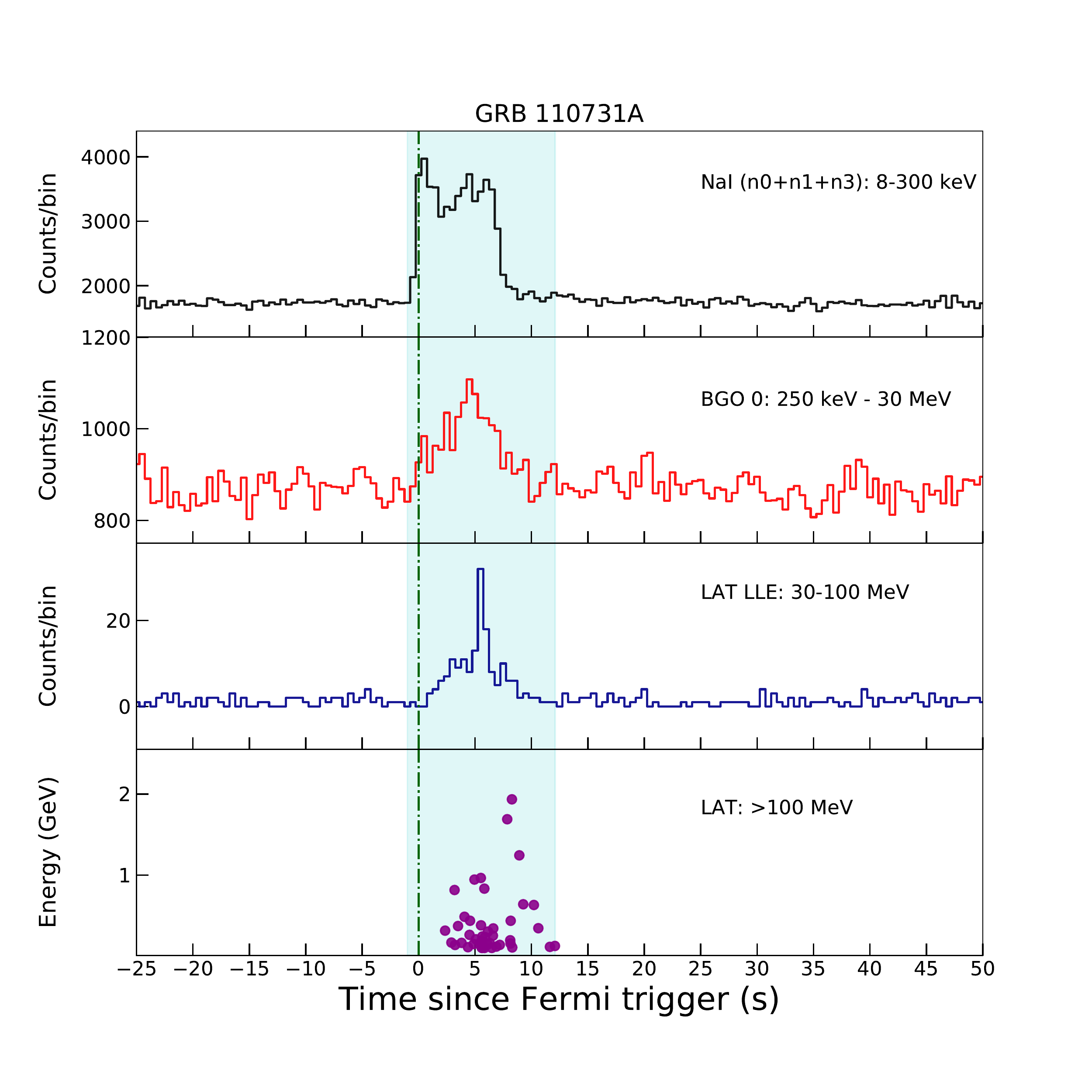}
    \includegraphics[width=.49\linewidth]{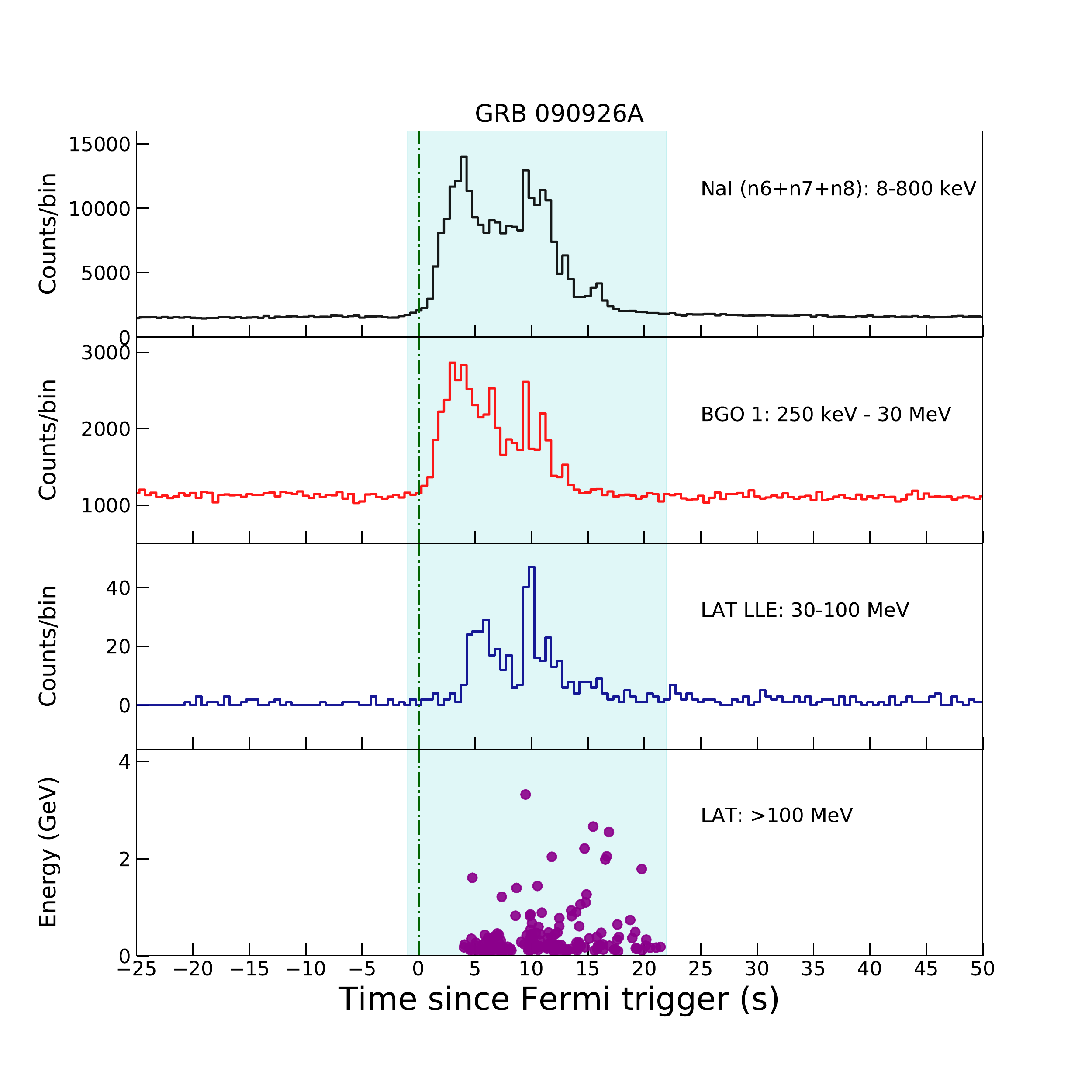}
    \includegraphics[width=.49\linewidth]{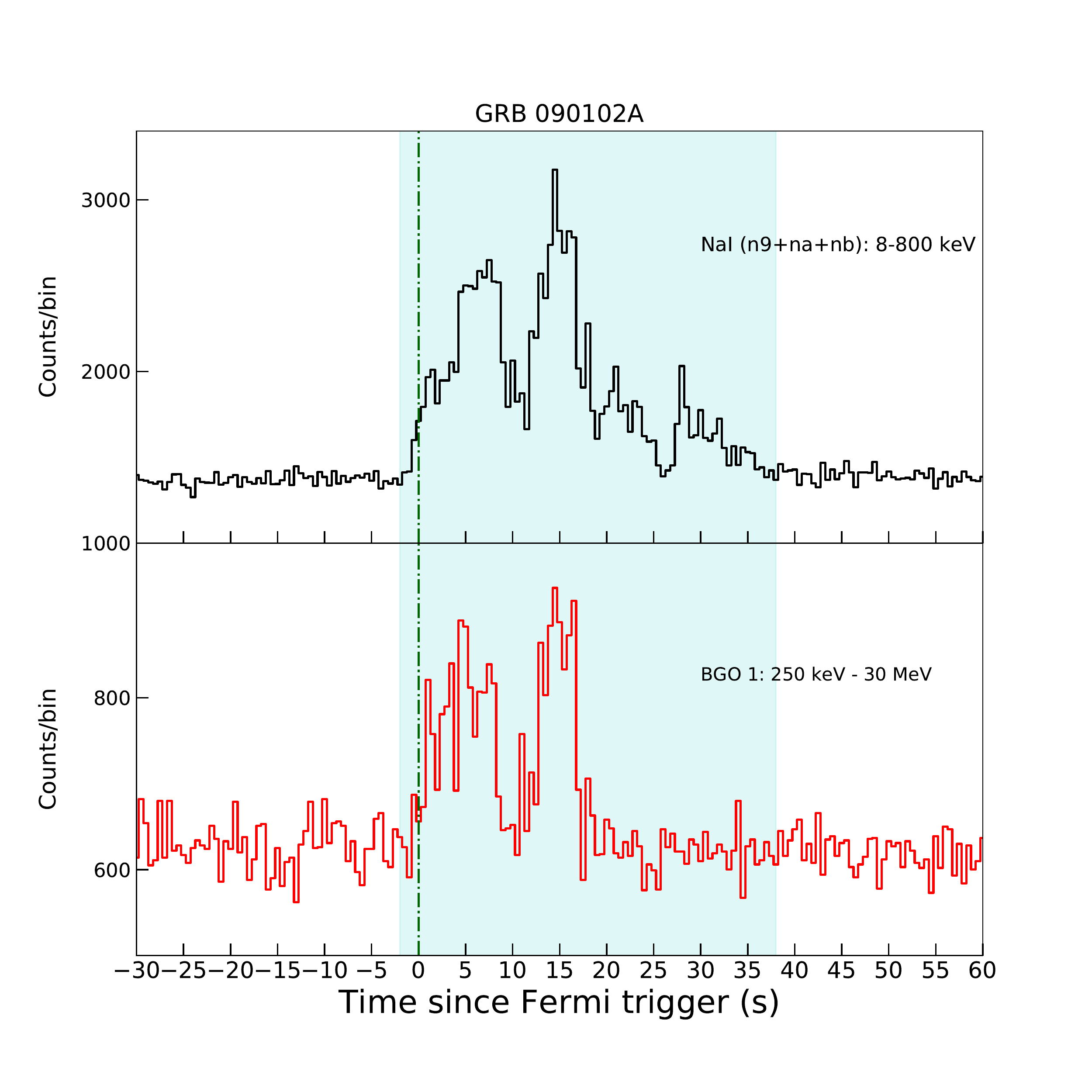}
    \caption{Figure \ref{lcs_1} continued.}
    \label{lcs_2}
\end{figure}

The prompt emission light curves of the eight GRBs as detected by Fermi GBM and LAT are presented in Figure \ref{lcs_1} and \ref{lcs_2}.
\label{fig:lc}

\section{Potential limitations and caveats}
\label{caveats}

\begin{figure}
    \centering
     \includegraphics[width=.90\linewidth]{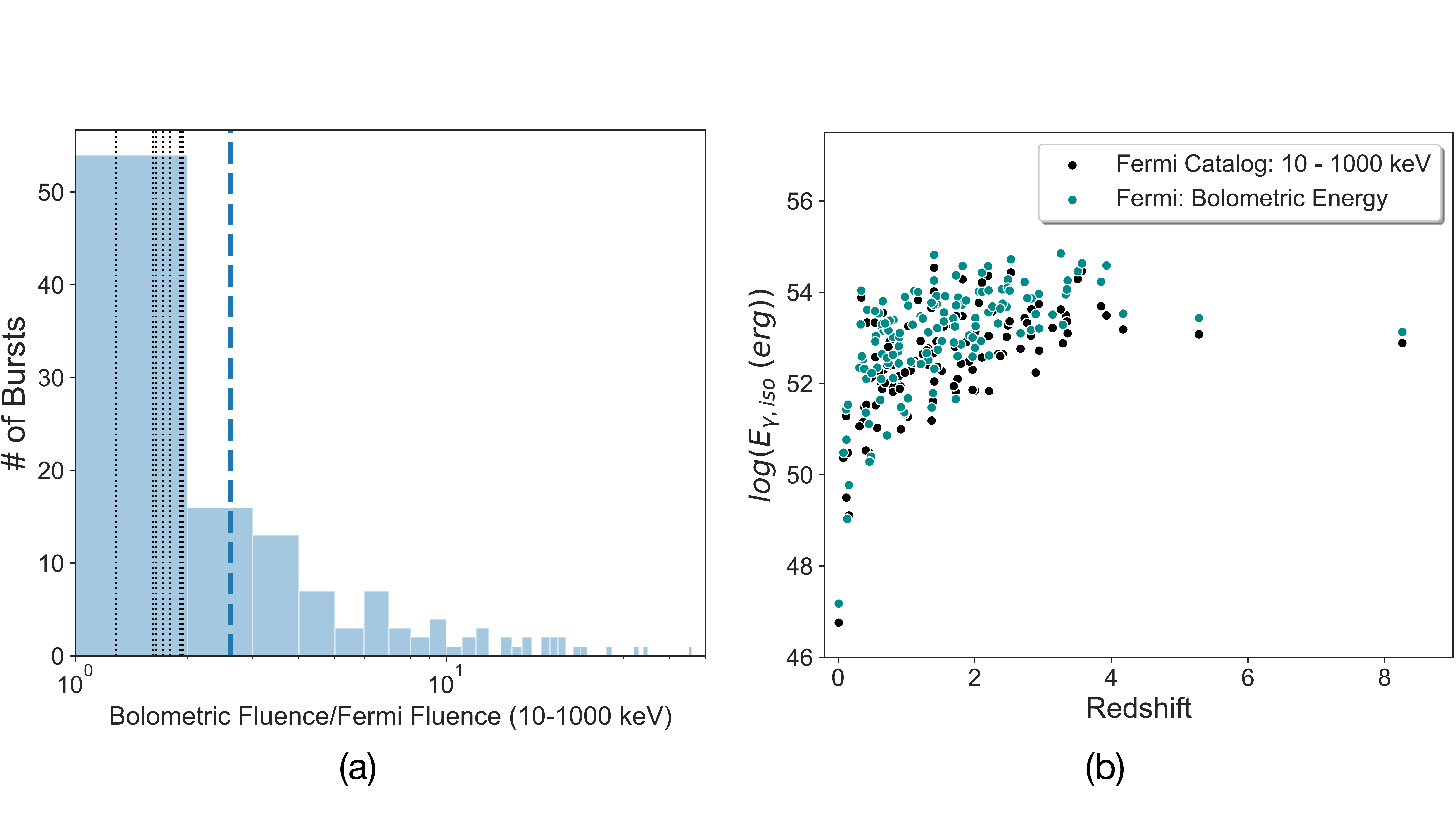}
    \caption{ (a) The distribution of the ratio of bolometric fluence to the {\it Fermi} catalog fluence is shown in the left panel. The median value of the distribution is $\sim 3$ and presented with blue dashed vertical line. The fluence ratio for the final 8 GRBs listed in the Table \ref{bh_sample} ranging between 1.3 and 2 are shown in black dotted vertical lines. (b) The isotropic burst energies estimated using the fluence in the energy range 10-1000 keV and bolometric fluence in the energy range 1 keV - 1 GeV or observed energy range (LAT-LLE GRBs observed above 1 GeV) are shown in black and teal circles respectively with respect to the measured redshift.}
    \label{fig:ratio_c}
\end{figure}

\begin{figure}
    \centering
     \includegraphics[width=.49\linewidth]{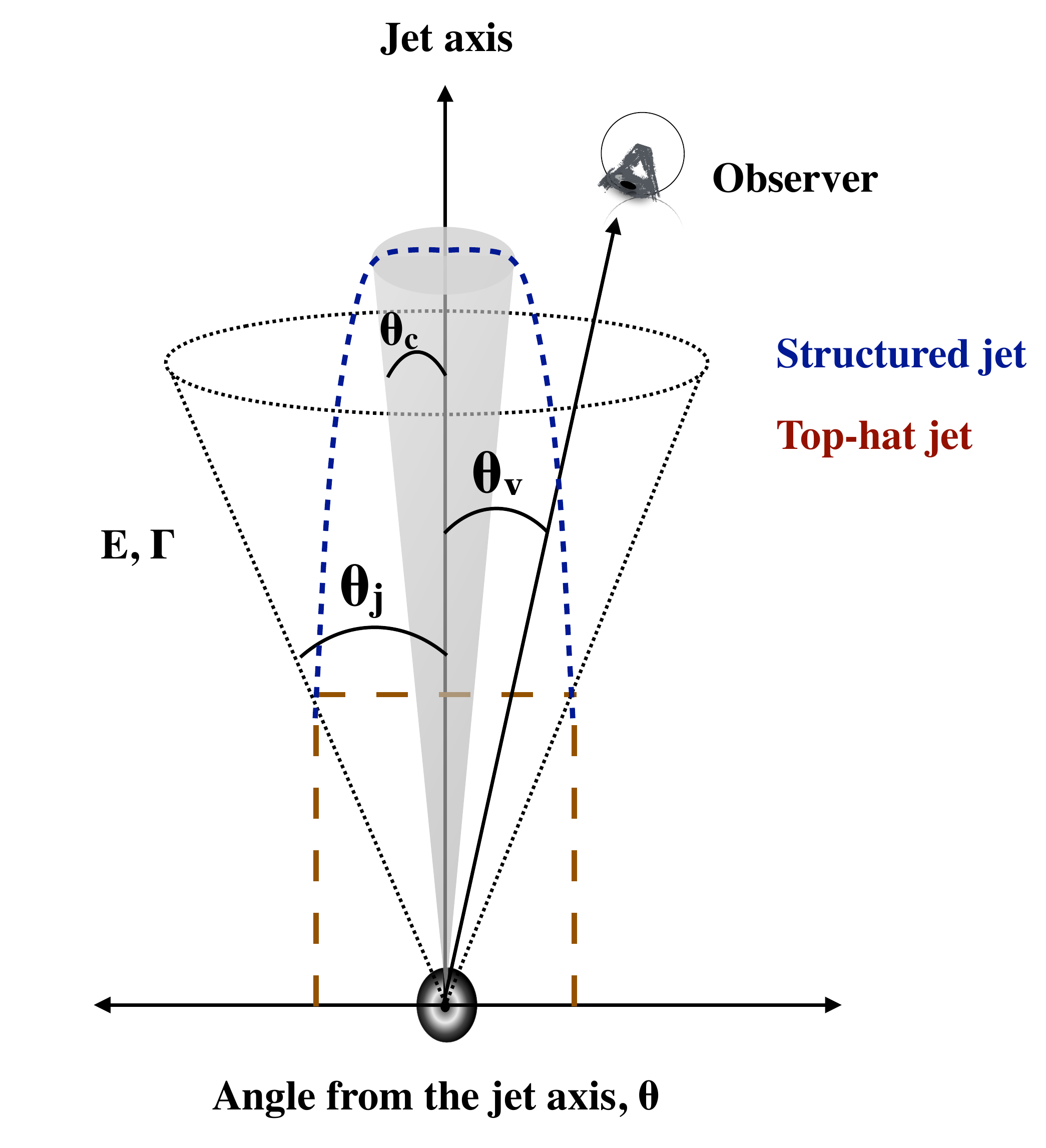}
    \caption{Above an illustration of a collimated GRB outflow with an opening angle, $\theta_j$ is shown. In case of a top-hat jet model, both the jet energy, $E$ and Lorentz factor, $\Gamma$ are angle independent (i.e constant, shown in brown long dashed line) within the $\theta_j$. On the other hand, in case of a structured jet model, the $E$ or $\Gamma$ remains a constant within the jet core, $\theta_c$ and beyond that possess a certain profile till $\theta_j$ (shown in blue short dashed line). $\theta_v$ is the viewing angle as measured from the jet axis.}
    \label{illustration_}
\end{figure}

 Below we discuss the different caveats that are involved in the estimation of the beam corrected total energy of the bursts, $E_{\gamma,beam}$:
\begin{itemize}
 \item Bolometric fluence:  In general, the burst fluence of {\it Fermi} detected GRBs are reported in the energy range $10-1000\, \rm keV$ which is a small energy window. Therefore, in order to assess the total energy of the GRB, we estimate the bolometric fluence by extrapolating the spectrum determined in the {\it Fermi} energy window (GBM: $8\, \rm keV$ - $40 \rm MeV$) to an energy interval of  $1 \, \rm keV$ - $1\, \rm GeV$. In case of GRBs with LAT detections with the highest energy photons greater than 1 GeV, the bolometric fluence is estimated for the energy interval until that highest energy value. We note that among the final 8 LAT detected long GRBs excepting GRB 120624B (till LLE, $100\; MeV$) and GRB 090102A (GBM only), the spectrum is well constrained until $\ge 1 \; GeV$ and therefore, their bolometric fluence estimates are robust.  
For the 135 GRBs with known redshifts in our study, the median of the ratio of bolometric fluence to the {\it Fermi} catalog fluence in the energy range, $10 -1000\, \rm keV$, is found to be nearly $3$. The ratio of the fluences is found to range between $1$ to $45$, as shown in the Figure \ref{fig:ratio_c}.  

\item Jet opening angle and viewing geometry: 
It has been shown by \citealt{VanEerten_etal_2010,Ryan_etal_2015} studies that it is likely to view the jet off-axis at a significant fraction of the jet opening angle. This can smear out the possible jet break and produce a smooth transition into the post jet break flux decay which is likely to be visible in a time span beyond the typical viewing window of Swift XRT ($10$ days). This may lead to over-estimation of the jet opening angle and thereby effect the estimated beam corrected energy of the GRB. This caveat is more relevant in cases where a clear achromatic jet break is not observed and where we have estimated a lower limit on the jet opening angle assuming on-axis viewing geometry.

\item Structured jet: The collimated GRB outflow can be in the form of
either a uniform (top-hat) or a structured jet \citep{granot2002off}. In case of a uniform jet, the total energy of the outflow is confined to a certain solid angle corresponding to an opening angle of $\theta_j$. On the other hand, in case of a structured jet, the jet energy or Lorentz factor remains a constant within the jet-core, $\theta_c$ and beyond that there exists a certain structure such as a decaying Lorentz factor or jet energy with respect to the angle measured from the jet axis (Figure \ref{illustration_})
until $\theta_j$. The GRB will be visible only when the line of sight of the observer is within the $\theta_j + 1/\Gamma(\theta_j)$. 

In case of a structured jet, for an off-axis observer when the viewing angle, $\theta_c < \theta_v < \theta_j$, the observed prompt energy flux of the burst is lower relative to the scenario when $\theta_v < \theta_c$. Thus, the $E_{\gamma,iso}$ estimate done in such a case will be a lower limit of the actual total burst energy. 
For on-axis observer, when these bursts have their jet core pointed 
towards the observer (i.e $\theta_v < \theta_c$). The prominent 
jet break observed in the late time afterglow emission can be then related to the scenario when $1/\Gamma (\theta_v\le \theta_c) = \theta_j$ \citep{Peng_etal_2005}. 
Since in a structured jet, the energy injection ($E$) and $\Gamma$ is angle dependent within the $\theta_j$, the  $E_{\gamma, iso}$ 
estimate as well as the beaming correction applied on the 
$E_{\gamma, iso}$ would give an overestimation of the true burst energy ($E_{\gamma,beam}$).  
However, to get a correct estimate of the burst energy, one should know the profile of the jet and the viewing angle. These estimates are generally difficult because of the degeneracy between various model parameters and currently out of the scope of this work. 
\end{itemize}

\section{Online Table}
\label{all_table}
The details of the 135 GRBs with known redshift and detected by Fermi during the years 2008-2019 are listed in the Tables \ref{sample_1}, \ref{sample_2}, \ref{sample_3} and \ref{sample_4}.
\begin{table}
\let\nobreakspace\relax
\begin{small}
	\caption{Properties of 135 GRBs}
	\label{sample_1}
	\begin{center}
	\hspace*{-2.5cm}
	\begin{tabular}{p{0.75cm}cccccccccc}
        \hline \hline
		S.No. & GRB name    & $T_{90}$ (\textit{Fermi})  & z    & LLE/LAT   & Fluence   & $E_{\gamma, iso}$   & $\theta_j$    & Reference & $E_{\gamma,beam}$ \\
		      &     & (s)                               &       &           & $erg/cm^2$      &   $log(E\;in\;erg)$   & $^{\circ}$    &       & $log(E\;in\;erg)$ \\
		\hline
    	\noalign{\vskip 1.5mm}
        1	&	191011A	&	25.09	&	1.722	&	NO	&	5.87E-07	&	51.66	&	--	&	--	&	--	\\	\noalign{\vskip 1.5mm}
        
        2	&	190829A	&	59.39	&	0.0785	&	NO	&	2.01E-05	&	50.48	&	--	&	--	&	--	\\	\noalign{\vskip 1.5mm}
        
        3	&	190719C	&	175.62	&	2.469	&	NO	&	3.29E-05	&	53.68	&	$3.2_{-0.5}^{+0.5}$	&	This work	&	50.88	\\	\noalign{\vskip 1.5mm}
        
        4	&	190114C	&	116.35	&	0.425	&	YES	&	8.45E-04	&	53.62	&	$>32.5$	&	\citealt{misra2019}	&	52.81	\\	\noalign{\vskip 1.5mm}
        
        5	&	181020A	&	15.10	&	2.938	&	NO	&	4.66E-05	&	53.96	&	$1.4_{-0.2}^{+0.2}$	&	This work	&	50.46	\\	\noalign{\vskip 1.5mm}
        
        6	&	181010A	&	9.73	&	1.39	&	NO	&	1.18E-06	&	51.79	&	--	&	--	&	--	\\	\noalign{\vskip 1.5mm}
        
        7	&	180728A	&	6.4	&	0.117	&	NO	&	7.91E-05	&	51.43	&	--	&	--	& --		\\	\noalign{\vskip 1.5mm}
        
        8	&	180720B	&	48.9	&	0.654	&	YES	&	5.34E-04	&	53.80	&	$>17.2$	&	This work	&	52.44	\\	\noalign{\vskip 1.5mm}
        
        9	&	180703A	&	20.74	&	0.6678	&	NO	&	1.15E-04	&	53.15	&	$>7.5$	&	This work	&	51.09	\\	\noalign{\vskip 1.5mm}
        
        10	&	180620B	&	46.72	&	1.1175	&	NO	&	3.10E-04	&	54.03	&	$5.9_{-0.9}^{+0.0}$	&	This work	&	51.76	\\	\noalign{\vskip 1.5mm}
        
        11	&	180314A	&	22.02	&	1.445	&	NO	&	9.87E-05	&	53.74	&	$>9.2$	&	This work	&	51.86	\\	\noalign{\vskip 1.5mm}
        
        12	&	180205A	&	15.36	&	1.409	&	NO	&	3.91E-06	&	52.32	& --		&	--	& --		\\	\noalign{\vskip 1.5mm}
        
        13	&	171222A	&	80.38	&	2.409	&	NO	&	4.01E-05	&	53.75	&	$>8.2$	&	This work	&	51.75	\\	\noalign{\vskip 1.5mm}
        
        14	&	171010A	&	107.27	&	0.3285	&	YES	&	6.85E-04	&	53.29	&	$>6$	&	\citealt{chand2019}	&	51.16	\\	\noalign{\vskip 1.5mm}
        
        15	&	170903A	&	25.6	&	0.886	&	NO	&	1.26E-05	&	52.44	&	--	&	--	&	--	\\	\noalign{\vskip 1.5mm}
        
        16	&	170817A	&	2.05	&	0.0093	&	NO	&	7.3E-07	&	47.18	&	--	& --		& --		\\	\noalign{\vskip 1.5mm}
        
        17	&	170705A	&	22.78	&	2.01	&	NO	&	2.64E-05	&	53.43	&	$>12.4$	&	This work	&	51.80	\\	\noalign{\vskip 1.5mm}
        
        18	&	170607A	&	20.93	&	0.632	&	NO	&	1.13E-05	&	52.10	& --		&	--	&	--	\\	\noalign{\vskip 1.5mm}
        
        19	&	170405A	&	78.59	&	3.51	&	YES	&	1.10E-04	&	54.46	&	$0.8_{-0.2}^{+0.2}$	&	This work	&	50.45	\\	\noalign{\vskip 1.5mm}
        
        20	&	170214A	&	122.88	&	2.53	&	YES	&	3.45E-04	&	54.72	&	$>3.7$	&	This work	&	52.04	\\	\noalign{\vskip 1.5mm}
        
        21	&	170113A	&	49.15	&	1.968	&	NO	&	3.93E-06	&	52.59	&	$12^{+1.9}_{-1.9}$	&	\citealt{li2018}	&	50.92	\\	\noalign{\vskip 1.5mm}
        
        22	&	161129A	&	36.1	&	0.645	&	NO	&	3.82E-05	&	52.64   &	$2.3_{-0.3}^{+0.3}$	&	This work	&	49.54	\\	\noalign{\vskip 1.5mm}
        
        23	&	161117A	&	122.18	&	1.549	&	NO	&	5.61E-05	&	53.55	&	$4.7_{-0.1}^{+0.1}$	&	\citealt{li2018}	&	51.08	\\	\noalign{\vskip 1.5mm}
        
        24	&	161017A	&	37.89	&	2.013	&	NO	&	1.75E-05	&	53.25	&	$2.8_{-0.5}^{+0.5}$	&	\citealt{tachibana2018}	&	50.34	\\	\noalign{\vskip 1.5mm}
        
        25	&	161014A	&	36.61	&	2.823	&	NO	&	8.1E-06 &	53.17	&	$2.2_{-0.3}^{+0.3}$	&	This work	&	50.04	\\	\noalign{\vskip 1.5mm}
        
        26	&	160821B	&	1.088	&	0.16	&	NO	&	9.11E-07	&	49.77	&	--	&	--	&	--	\\	\noalign{\vskip 1.5mm}
        
        27	&	160804A	&	131.59	&	0.736	&	NO	&	1.86E-05	&	52.45	& --		& --		& --		\\	\noalign{\vskip 1.5mm}
        
        28	&	160629A	&	64.77	&	3.332	&	NO	&	3.73E-05	&	53.95	&	$>2.9$	&	This work	&	51.06	\\	\noalign{\vskip 1.5mm}
        
        29	&	160625B	&	453.38	&	1.406	&	YES	&	1.24E-03	&	54.82	&	$3.6_{-0.2}^{+0.2}$	&	\citealt{alexander2017}	&	52.11	\\	\noalign{\vskip 1.5mm}
        
        30	&	160624A	&	0.38	&	0.483	&	NO	&	3.92E-07	&	50.40	&	--	& --		& --		\\	\noalign{\vskip 1.5mm}
        
        31	&	160623A	&	107.78	&	0.367	&	YES	&	9.24E-05	&	52.53	&	$13_{-2.8}^{+2.8}$	&	\citealt{chen2020}	&	50.93	\\	\noalign{\vskip 1.5mm}
        
        32	&	160509A	&	369.67	&	1.17	&	YES	&	2.69E-04	&	54.00	&	$3.9_{-0.2}^{+0.2}$	&	\citealt{laskar2016}	&	51.37	\\	\noalign{\vskip 1.5mm}
        
        33	&	151027A	&	123.39	&	0.81	&	NO	&	1.36E-04	&	53.39	&	$7.7_{-0.2}^{+0.2}$	&	\citealt{li2018}	&	51.35	\\	\noalign{\vskip 1.5mm}
        
        34	&	150821A	&	103.43	&	0.755	&	NO	&	1.5E-04	&	53.37	&	$0.7_{-0.2}^{+0.2}$	&	This work	&	49.27	\\	\noalign{\vskip 1.5mm}
        
        35	&	150727A	&	49.41	&	0.313	&	NO	&	8.49E-05	&	52.34	&	--	& --		& --		\\	\noalign{\vskip 1.5mm}

        36	&	150514A	&	10.81	&	0.807	&	YES	&	7.22E-06	&	52.17	&	--	& --		&	--	\\	\noalign{\vskip 1.5mm}
         \hline \\
            \end{tabular}
    \vspace{-0.5cm}
\end{center}
\end{small}
\end{table}

\begin{table}
\let\nobreakspace\relax
\begin{small}
	\caption{Properties of 135 GRBs (continued)}
	\label{sample_2}
	\begin{center}
	\hspace*{-2.5cm}
	\begin{tabular}{p{0.75cm}cccccccccc}
        \hline \hline
		S.No. & GRB name    & $T_{90}$ (\textit{Fermi})  & z    & LLE/LAT   & Fluence   & $E_{\gamma, iso}$   & $\theta_j$    & Reference & $E_{\gamma,beam}$ \\
		      &     & (s)                               &       &           &    $erg/cm^2$       &   $log(E\;in\;erg)$   & $^{\circ}$    &       & $log(E\;in\;erg)$ \\
		\hline
        37	&	150403A	&	22.27	&	2.06	&	YES	&	9.57E-05	&	54.01	&	7.74	&	\citealt{pisani2016}	&	51.97	\\	\noalign{\vskip 1.5mm}
        
        38	&	150314A	&	10.69	&	1.758	&	YES	&	9.48E-05	&	53.88	&	$0.7_{-0.1}^{+0.1}$	&	This work	&	49.79	\\	\noalign{\vskip 1.5mm}
        
        39	&	150301B	&	13.31	&	1.5169	&	NO	&	1.38E-05	&	52.93	&	$>4.3$	&	This work	&	50.38	\\	\noalign{\vskip 1.5mm}
        
        40	&	150120A	&	3.33	&	0.46	&	NO	&	3.35E-07	&	50.28	& --		& --		& --		\\	\noalign{\vskip 1.5mm}
        
        41	&	150101B	&	0.08	&	0.134	&	NO	&	2.38E-07	&	49.03	& --		& --	& --		\\	\noalign{\vskip 1.5mm}
        
        42	&	141225A	&	56.32	&	0.915	&	NO	&	5.78E-05	&	53.13	&	$>4.9$	&	This work	&	50.69	\\	\noalign{\vskip 1.5mm}
        
        43	&	141221A	&	23.81	&	1.452	&	NO	&	2.9E-05	&	53.21	&	$>5.8$	&	This work	&	50.92	\\	\noalign{\vskip 1.5mm}
        
        44	&	141220A	&	7.67	&	1.3195	&	NO	&	6.93E-06	&	52.51	&	$>6.6$	&	This work	&	50.34	\\	\noalign{\vskip 1.5mm}
        
        45	&	141028A	&	31.49	&	1.82	&	YES	&	6.26E-05	&	53.73	&	$>7.5$	&	This work	&	51.66	\\	\noalign{\vskip 1.5mm}
        
        46	&	141004A	&	2.56	&	0.573	&	NO	&	2.46E-05	&	52.35	&	--	& --		& --		\\	\noalign{\vskip 1.5mm}
        
        47	&	140907A	&	35.84	&	1.21	&	NO	&	6.64E-06	&	52.42	& --		&	--	&	--	\\	\noalign{\vskip 1.5mm}
        
        48	&	140808A	&	4.48	&	3.29	&	NO	&	8.17E-06	&	53.28. &	$>5.1$	&	This work	&	50.89 \\	\noalign{\vskip 1.5mm}
        
        49	&	140801A	&	7.17	&	1.32	&	NO	&	2.8E-05	&	53.12	&	$>7.3$	&	This work	&	51.03	\\	\noalign{\vskip 1.5mm}
        
        50	&	140703A	&	83.97	&	3.14	&	NO	&	1.48E-05	&	53.51	&	$8.9_{-0.1}^{+0.1}$	&	\citealt{li2018}	&	51.59	\\	\noalign{\vskip 1.5mm}
        
        51	&	140623A	&	111.10	&	1.92	&	NO	&	1.2E-05	&	53.05	&	$>9.1$	&	This work	&	51.16	\\	\noalign{\vskip 1.5mm}
        
        52	&	140620A	&	45.82	&	0.88	&	NO	&	2.36E-05	&	52.70	&	$>11.7$	&	This work	&	51.02	\\	\noalign{\vskip 1.5mm}
        
        53	&	140606B	&	22.78	&	0.384	&	NO	&	5.3E-05	&	52.32	& --		& --		& --		\\	\noalign{\vskip 1.5mm}
        
        54	&	140512A	&	147.97	&	0.725	&	NO	&	1.0E-04	&	53.17	&	$6_{-0.1}^{+0.1}$	&	\citealt{li2018}	&	50.90	\\	\noalign{\vskip 1.5mm}
        
        55	&	140508A	&	44.29	&	1.027	&	NO	&	1.7E-04	&	53.71	&	$>9.9$	&	This work	&	51.88	\\	\noalign{\vskip 1.5mm}
        
        56	&	140506A	&	64.13	&	0.889	&	NO	&	2.98E-05	&	52.82	&	$>18.8$	&	This work	&	51.54	\\	\noalign{\vskip 1.5mm}
        
        57	&	140423A	&	95.23	&	3.26	&	NO	&	3.04E-04	&	54.85	&	$>4.1$	&	This work	&	52.25	\\	\noalign{\vskip 1.5mm}
        
        58	&	140304A	&	31.23	&	5.283	&	NO	&	5.51E-06	&	53.43	&	$1.2_{-0.3}^{+0.3}$	&	This work	&	49.74	\\	\noalign{\vskip 1.5mm}
        
        59	&	140213A	&	18.62	&	1.2076	&	NO	&	7.4E-05	&	53.47	&	$4.6_{-1.3}^{+1.3}$	&	This work	&	50.99	\\	\noalign{\vskip 1.5mm}
        
        60	&	140206A	&	27.26	&	2.73	&	NO	&	9.65E-05	&	54.22	&	$2.9_{-0.5}^{+0.5}$	&	This work	&	51.33	\\	\noalign{\vskip 1.5mm}
        
        61	&	131231A	&	31.23	&	0.642	&	YES	&	1.75E-04	&	53.30	&	$>8.7$	&	This work	&	51.36	\\	\noalign{\vskip 1.5mm}
        
        62	&	131108A	&	18.18	&	2.4	&	YES	&	8.33E-05	&	54.06	&	$>5.2$	&	This work	&	51.68	\\	\noalign{\vskip 1.5mm}
        
        63	&	131105A	&	112.64	&	1.686	&	NO	&	3.55E-05	&	53.42	&	$4.3_{-0.1}^{+0.1}$	&	\citealt{li2018}	&	50.86	\\	\noalign{\vskip 1.5mm}
        
        64	&	131011A	&	77.06	&	1.874	&	NO	&	7.31E-05	&	53.82	&	$>3.5$	&	This work	&	51.09	\\	\noalign{\vskip 1.5mm}
        
        65	&	131004A	&	1.15	&	0.717	&	NO	&	5.10E-07	&	50.86 &	--	&	--	& --		\\	\noalign{\vskip 1.5mm}
        
        66	&	130925A	&	215.56	&	0.347	&	NO	&	1.22E-04	&	52.59	&	$>33$	&	This work	&	52.01	\\	\noalign{\vskip 1.5mm}
        
        67	&	130702A	&	58.88	&	0.145	&	YES	&	1.28E-04	&	51.54	&	--	& --		& --		\\	\noalign{\vskip 1.5mm}
        
        68	&	130612A	&	7.42	&	2.006	&	NO	&	5.98E-06	&	52.78	&	$11.4_{-0.4}^{+0.4}$	&	\citealt{li2018}	&	51.08	\\	\noalign{\vskip 1.5mm}
        
        69	&	130610A	&	21.76	&	2.092	&	NO	&	7.72E-06	&	52.93	&	$>6.9$	&	This work	&	50.79	\\	\noalign{\vskip 1.5mm}
        
        70	&	130518A	&	48.58	&	2.488	&	YES	&	1.28E-04	&	54.28	&	$>4.9$	&	This work	&	51.84	\\	\noalign{\vskip 1.5mm}
        
        71	&	130427A	&	138.24	&	0.3399	&	YES	&	3.51E-03	&	54.03	&	$>5$	&	\citealt{perley2014}	&	$<52$	\\	\noalign{\vskip 1.5mm}
        
        72	&	130420A	&	104.96	&	1.297	&	NO	&	1.01E-05	&	52.66	&	$15.1_{-5.1}^{+5.1}$	&	\citealt{li2018}	&	51.23	\\	\noalign{\vskip 1.5mm}
        
                \hline \\
    \end{tabular}
    \vspace{-0.5cm}
\end{center}
\end{small}
\end{table}
 
 \begin{table}
\let\nobreakspace\relax
\begin{small}
	\caption{Properties of 135 GRBs (continued)}
	\label{sample_3}
	\begin{center}
	\hspace*{-2.5cm}
	\begin{tabular}{p{0.75cm}cccccccccc}
        \hline \hline
		S.No. & GRB name    & $T_{90}$ (\textit{Fermi})  & z    & LLE/LAT   & Fluence   & $E_{\gamma, iso}$   & $\theta_j$    & Reference & $E_{\gamma,beam}$ \\
		      &     & (s)                               &       &           &    $erg/cm^2$       &   $log(E\;in\;erg)$   & $^{\circ}$    &       & $log(E\;in\;erg)$ \\
		\hline       
        73	&	130215A	&	143.75	&	0.597	&	NO	&	3.52E-04	&	53.54	&	--	&	No XRT	&	--	\\	\noalign{\vskip 1.5mm}
        
        74	&	121211A	&	5.63	&	1.023	&	NO	&	1.64E-06	&	51.68	&	--	& --		& --		\\	\noalign{\vskip 1.5mm}
        
        75	&	121128A	&	17.34	&	2.2	&	NO	&	3.02E-05	&	53.56	&	$10.5_{-0.3}^{+0.3}$	&	\citealt{li2018}	&	51.78	\\	\noalign{\vskip 1.5mm}
        
        76	&	120909A	&	112.07	&	3.93	&	NO	&	1.23E-04	&	54.58	&	$>3.2$	&	This work	&	51.79	\\	\noalign{\vskip 1.5mm}
        
        77	&	120907A	&	5.76	&	0.97	&	NO	&	8.91E-07	&	51.36	& --		& --		& --		\\	\noalign{\vskip 1.5mm}
        
        78	&	120811C	&	14.34	&	2.671	&	NO	&	7.5E-06	&	53.10	&	$6.4_{-0.02}^{+0.02}$	&	\citealt{li2018}	&	50.89	\\	\noalign{\vskip 1.5mm}
        
        79	&	120729A	&	25.47	&	0.8	&	YES	&	5.86E-05	&	53.02	&	$1.2_{-0.3}^{+0.3}$	&	\citealt{wang2018}	&	49.34	\\	\noalign{\vskip 1.5mm}
        
        80	&	120716A	&	226.05	&	2.486	&	NO	&	8.47E-05	&	54.10	&	$>4.2$	&	This work	&	51.53	\\	\noalign{\vskip 1.5mm}
        
        81	&	120712A	&	22.53	&	4.1745	&	NO	&	9.83E-06	&	53.53	&	$5_{-0.02}^{+0.02}$	&	\citealt{li2018}	&	51.10	\\	\noalign{\vskip 1.5mm}
        
        82	&	120711A	&	44.03	&	1.405	&	YES	&	3.39E-04	&	54.26	&	$>4.6$	&	This work	&	51.77	\\	\noalign{\vskip 1.5mm}
        
        83	&	120624B	&	271.36	&	2.1974	&	YES	&	3.11E-04	&	54.57	&	$>5.9$	&	This work	&	52.28	\\	\noalign{\vskip 1.5mm}
        
        84	&	120326A	&	11.78	&	1.798	&	NO	&	8.57E-06	&	52.86	&	$4.6_{-0.2}^{+0.2}$	&	\citealt{song2016}	&	50.36	\\	\noalign{\vskip 1.5mm}
        
        85	&	120119A	&	55.3	&	1.728	&	NO	&	6.59E-05	&	53.71	&	$1.8_{-0.1}^{+01}$	&	\citealt{song2018}	&	50.42	\\	\noalign{\vskip 1.5mm}
        
        86	&	120118B	&	37.82	&	2.943	&	NO	&	8.18E-06	&	53.20	&	$>8$	&	This work	&	51.20	\\	\noalign{\vskip 1.5mm}
        
        87	&	111228A	&	99.84	&	0.714	&	NO	&	2.58E-05	&	52.56	&	$7.3_{-0.2}^{+0.2}$	&	\citealt{li2018}	&	50.47	\\	\noalign{\vskip 1.5mm}
        
        88	&	111117A	&	0.43	&	2.211	&	NO	&	3.39E-06	&	52.61	&	6	&	\citealt{song2018}	&	50.35	\\	\noalign{\vskip 1.5mm}
        
        89	&	111107A	&	12.03	&	2.893	&	NO	&	1.74E-05	&	53.52	&	$>5.4$	&	This work	&	51.17	\\	\noalign{\vskip 1.5mm}
        
        90	&	110818A	&	67.07	&	3.36	&	NO	&	7.36E-05	&	54.25	&	$1.8_{-0.3}^{+0.3}$	&	This work	&	50.93	\\	\noalign{\vskip 1.5mm}
        
        91	&	110731A	&	7.48	&	2.83	&	YES	&	3.95E-05	&	53.86	&	$28.9_{-0.7}^{+0.0}$	&	\citealt{zhang2015}	&	52.96	\\	\noalign{\vskip 1.5mm}
        
        92	&	110213A	&	34.31	&	1.46	&	NO	&	9.65E-06	&	52.74	&	8.1	&	s\citealt{song2018}	&	50.75	\\	\noalign{\vskip 1.5mm}
        
        93	&	110128A	&	7.94	&	2.339	&	NO	&	1.56E-05	&	53.32	&	$>8.9$	&	This work	&	51.34	\\	\noalign{\vskip 1.5mm}
        
        94	&	110106B	&	35.52	&	0.618	&	NO	&	4.11E-06	&	51.64	&	--	& --		&	--	\\	\noalign{\vskip 1.5mm}
        
        95	&	101219B	&	51.01	&	0.5519	&	NO	&	1.30E-04	&	53.04	&	17.1	&	\citealt{song2019}	&	51.68	\\	\noalign{\vskip 1.5mm}
        
        96	&	101213A	&	45.06	&	0.414	&	NO	&	2.73E-05	&	52.10	&	--	& --		&	 --	\\	\noalign{\vskip 1.5mm}
        
        97	&	100906A	&	110.59	&	1.727	&	NO	&	2.99E-04	&	54.37	&	$4.4_{-0.1}^{+0.1}$	&	\citealt{li2018}	&	51.84	\\	\noalign{\vskip 1.5mm}
        
        98	&	100816A	&	2.04	&	0.8035	&	NO	&	2.37E-05	&	52.63	&	$28.2_{-3.7}^{+0.02}$	&	\citealt{zhang2015}	&	51.71	\\	\noalign{\vskip 1.5mm}
        
        99	&	100814A	&	150.53	&	1.44	&	NO	&	1.47E-04	&	53.91	&	$5.2_{-0.1}^{+0.1}$	&	\citealt{li2018}	&	51.53	\\	\noalign{\vskip 1.5mm}
        
        100	&	100728B	&	10.24	&	2.106	&	NO	&	9.18E-05	&	54.01	&	3.6	&	\citealt{song2018}	&	51.30	\\	\noalign{\vskip 1.5mm}
        
        101	&	100728A	&	165.38	&	1.567	&	YES	&	1.26E-04	&	53.91	&	$1.6_{-0.3}^{+0.3}$	&	This work	&	50.51	\\	\noalign{\vskip 1.5mm}
        
        102	&	100625A	&	0.24	&	0.452	&	NO	&	2.33E-06	&	51.11	&	--	&	--	&	--	\\	\noalign{\vskip 1.5mm}
        
        103	&	100615A	&	37.38	&	1.398	&	NO	&	8.7234E-06	&	52.66	&	$25.6_{-2.2}^{+2.2}$	&	\citealt{zhang2015}	&	51.65	\\	\noalign{\vskip 1.5mm}
        
        104	&	100414A	&	26.5	&	1.368	&	YES	&	1.17E-04	&	53.77	&	$>8.2$	&	This work	&	51.78	\\	\noalign{\vskip 1.5mm}
        
        105	&	100206A	&	0.18	&	0.4068	&	NO	&	5.12E-06	&	51.36	&	--	& --		&	--	\\	\noalign{\vskip 1.5mm}
        
        106	&	100117A	&	0.26	&	0.92	&	NO	&	1.29E-06	&	51.48	&	--	&	--	&	--	\\	\noalign{\vskip 1.5mm}
        
        107	&	091208B	&	12.48	&	1.063	&	YES	&	9.70E-06	&	52.48	&	7.3	&	\citealt{nemmen2012}	&	50.39	\\	\noalign{\vskip 1.5mm}
        
        108	&	091127	&	8.7	&	0.49	&	NO	&	2.57E-05	&	52.23	&	--	&	--	&	--	\\	\noalign{\vskip 1.5mm}
        
                        \hline \\
    \end{tabular}
    \vspace{-0.5cm}
\end{center}
\end{small}
\end{table}
 
 \begin{table}
\let\nobreakspace\relax
\begin{small}
	\caption{Properties of 135 GRBs (continued)}
	\label{sample_4}
	\begin{center}
	\hspace*{-2.5cm}
	\begin{tabular}{p{0.75cm}cccccccccc}
        \hline \hline
		S.No. & GRB name    & $T_{90}$ (\textit{Fermi})  & z    & LLE/LAT   & Fluence   & $E_{\gamma, iso}$   & $\theta_j$    & Reference & $E_{\gamma,beam}$ \\
		      &     & (s)                               &       &           &      $erg/cm^2$     &   $log(E\;in\;erg)$   & $^{\circ}$    &       & $log(E\;in\;erg)$ \\
		\hline       
        
        109	&	091024	&	93.95	&	1.092	&	NO	&	5.91E-05	&	53.29	&	4.07	&	\citealt{song2018}	&	50.69	\\	\noalign{\vskip 1.5mm}
        
        110	&	091020	&	24.26	&	1.71	&	NO	&	2.31E-05	&	53.25	&	6.9	&	\citealt{nemmen2012}	&	51.11	\\	\noalign{\vskip 1.5mm}
        
        111	&	 091003A	&	20.22	&	0.8969	&	YES	&	4.63E-05	&	53.01	&	$>14.1$	&	This work	&	51.50	\\	\noalign{\vskip 1.5mm}
        
        112	&	090927	&	0.51	&	1.37	&	NO	&	5.84E-07	&	51.47	&	--	& --		&	--	\\	\noalign{\vskip 1.5mm}
        
        113	&	090926B	&	64.0	&	1.24	&	NO	&	6.31E-05	&	53.42	&	$0.4_{-0.1}^{+0.1}$	&	This work	&	48.73   \\	\noalign{\vskip 1.5mm}
        
        114	&	090926	&	13.76	&	2.1062	&	YES	&	2.41E-04	&	54.43	&	$9_{-2}^{+4}$	&	\citealt{cenko2011}	&	52.52	\\	\noalign{\vskip 1.5mm}
        
        115	&	090902B	&	19.33	&	1.822	&	YES	&	4.37E-04	&	54.57	&	$3.9_{-0.2}^{+0.2}$	&	\citealt{cenko2011}	&	51.94	\\	\noalign{\vskip 1.5mm}
        
        116	&	090618	&	112.39	&	0.54	&	NO	&	4.81E-04	&	53.59	&	6.7	&	\citealt{nemmen2012}	&	51.42	\\	\noalign{\vskip 1.5mm}
        
        117	&	090516	&	123.14	&	3.85	&	NO	&	5.59E-05	&	54.23	&	$3.5_{-0.1}^{+0.1}$	& \citealt{li2018}	&	51.50	\\	\noalign{\vskip 1.5mm}
        
        118	&	090510	&	0.96	&	0.903	&	YES	&	5.64E-05	&	53.11	&	$14.1_{-0.1}^{+0.1}$	&	\citealt{li2018}	&	51.58	\\	\noalign{\vskip 1.5mm}
        
        119	&	090424	&	14.14	&	0.544	&	NO	&	1.03E-04	&	52.92	&	6.7	&	\citealt{nemmen2012}	&	50.76	\\	\noalign{\vskip 1.5mm}
        
        120	&	090423	&	7.17	&	8.26	&	NO	&	1.43E-06	&	53.13	&	$22.5_{-15.1}^{+0.6}$	&	\citealt{zhang2015}	&	52.01	\\	\noalign{\vskip 1.5mm}
        
        121	&	090328	&	61.7	&	0.736	&	YES	&	9.66E-05	&	53.16	&	$4.2_{-0.8}^{+1.3}$	&	\citealt{cenko2011}	&	50.59	\\	\noalign{\vskip 1.5mm}
        
        122	&	090323	&	133.89	&	3.57	&	YES	&	1.60E-04	&	54.63	&	$2.8_{-0.1}^{+0.4}$	&	\citealt{cenko2011}	&	51.71	\\	\noalign{\vskip 1.5mm}
        
        123	&	090113	&	17.41	&	1.7493	&	NO	&	4.95E-06	&	52.60	&	$6.9_{-4.1}^{+7.8}$	&	\citealt{zhang2015}	&	50.46	\\	\noalign{\vskip 1.5mm}
        
        124	&	090102	&	26.62	&	1.547	&	YES	&	3.59E-05	&	53.36	&	$23.9_{-12.1}^{+1.1}$	&	\citealt{zhang2015}	&	52.29	\\	\noalign{\vskip 1.5mm}
        
        125	&	081222	&	18.88	&	2.77	&	NO	&	4.17E-05	&	53.87	&	2.8	&	\citealt{nemmen2012}	&	50.94	\\	\noalign{\vskip 1.5mm}
        
        126	&	081221	&	29.7	&	2.26	&	NO	&	3.86E-05	&	53.68	&	$4.3_{-0.1}^{+0.1}$	&	\citealt{li2018}	&	51.12	\\	\noalign{\vskip 1.5mm}
        
        127	&	081121	&	41.98	&	2.512	&	NO	&	7.19E-05	&	54.03	&	$>7.4$	&	This work	&	51.95	\\	\noalign{\vskip 1.5mm}
        
        128	&	081109	&	58.37	&	0.9787	&	NO	&	2.99E-04	&	53.90	&	$>7.5$	&	This work	&	51.83	\\	\noalign{\vskip 1.5mm}
        
        129	&	081008	&	126.72	&	1.9685	&	NO	&	5.88E-05	&	53.01	&	$6.1_{-0.3}^{+0.3}$	&	\citealt{li2018}	&	50.76	\\	\noalign{\vskip 1.5mm}
        
        130	&	080928	&	14.34	&	1.692	&	NO	&	1.06E-05	&	52.90	&	$2.4_{-0.4}^{+0.4}$	&	This work	&	49.82	\\	\noalign{\vskip 1.5mm}
        
        131	&	080916A	&	46.34	&	0.689	&	NO	&	1.58E-04	&	53.32	&	$>14.9$	&	This work	&	51.84	\\	\noalign{\vskip 1.5mm}
        
        132	&	080905B	&	105.98	&	2.374	&	NO	&	3.2E-05	&	53.64	&	--	& --		& --		\\	\noalign{\vskip 1.5mm}
        
        133	&	080905A	&	0.96	&	0.1218	&	NO	&	1.58E-05	&	50.77	&	$6.7_{-0.2}^{+0.2}$	&	\citealt{li2018}	&	51.48	\\	\noalign{\vskip 1.5mm}
        
        134	&	080810	&	75.20	&	3.35	&	NO	&	4.76E-05	&	54.06	&	3.83	&	\citealt{song2018}	&	51.41	\\	\noalign{\vskip 1.5mm}
        
        135	&	080804	&	24.70	&	2.2045	&	NO	&	9.11E-05	&	54.04	&	$2.9_{-0.8}^{+0.8}$	&	\citealt{wang2018}	&	51.16	\\	\noalign{\vskip 1.5mm}

        \hline \\
    \end{tabular}
    \vspace{-0.5cm}
\end{center}
\end{small}
\end{table}


\bibliography{ref_bh}

\begin{thebibliography}{}
\expandafter\ifx\csname natexlab\endcsname\relax\def\natexlab#1{#1}\fi
\providecommand{\url}[1]{\href{#1}{#1}}
\providecommand{\dodoi}[1]{doi:~\href{http://doi.org/#1}{\nolinkurl{#1}}}
\providecommand{\doeprint}[1]{\href{http://ascl.net/#1}{\nolinkurl{http://ascl.net/#1}}}
\providecommand{\doarXiv}[1]{\href{https://arxiv.org/abs/#1}{\nolinkurl{https://arxiv.org/abs/#1}}}

\bibitem[{{Abbott} {et~al.}(2019){Abbott}, {Abbott}, {Abbott}, {Abraham},
  {Acernese}, {Ackley}, {Adams}, {Adhikari}, {Adya}, {Affeldt}, {Agathos},
  {Agatsuma}, {Aggarwal}, {Aguiar}, {Aiello}, {Ain}, {Ajith}, {Allen},
  {Allocca}, {Aloy}, {Altin}, {Amato}, {Ananyeva}, {Anderson}, {Anderson},
  {Angelova}, {Antier}, {Appert}, {Arai}, {Araya}, {Areeda}, {Ar{\`e}ne},
  {Arnaud}, {Arun}, {Ascenzi}, {Ashton}, {Aston}, {Astone}, {Aubin}, {Aufmuth},
  {AultONeal}, {Austin}, {Avendano}, {Avila-Alvarez}, {Babak}, {Bacon},
  {Badaracco}, {Bader}, {Bae}, {Baker}, {LIGO Scientific Collaboration}, \&
  {Virgo Collaboration}}]{Abbott_etal_2019}
{Abbott}, B.~P., {Abbott}, R., {Abbott}, T.~D., {et~al.} 2019, \apjl, 882, L24,
  \dodoi{10.3847/2041-8213/ab3800}

\bibitem[{Abbott {et~al.}(2020)Abbott, Abbott, Abraham, Acernese, Ackley,
  Adams, Adhikari, Adya, Affeldt, Agathos, others, {LIGO Scientific
  Collaboration}, \& {Virgo Collaboration}}]{Abbott2020}
Abbott, R., Abbott, T., Abraham, S., {et~al.} 2020, \apjl, 896, L44,
  \dodoi{10.3847/2041-8213/ab960f}

\bibitem[{Aghanim {et~al.}(2018)Aghanim, Akrami, Ashdown, Aumont, Baccigalupi,
  Ballardini, Banday, Barreiro, Bartolo, Basak, {et~al.}}]{2018_planck}
Aghanim, N., Akrami, Y., Ashdown, M., {et~al.} 2018, arXiv preprint
  arXiv:1807.06209

\bibitem[{{Ajello} {et~al.}(2019){Ajello}, {Arimoto}, {Axelsson}, {Baldini},
  {Barbiellini}, {Bastieri}, {Bellazzini}, {Bhat}, {Bissaldi}, {Blandford},
  {Bonino}, {Bonnell}, {Bottacini}, {Bregeon}, {Bruel}, {Buehler}, {Cameron},
  {Caputo}, {Caraveo}, {Cavazzuti}, {Chen}, {Cheung}, {Chiaro}, {Ciprini},
  {Costantin}, {Crnogorcevic}, {Cutini}, {Dainotti}, {D'Ammand o}, {de la Torre
  Luque}, {de Palma}, {Desai}, {Desiante}, {Di Lalla}, {Di Venere}, {Fana
  Dirirsa}, {Fegan}, {Franckowiak}, {Fukazawa}, {Funk}, {Fusco}, {Gargano},
  {Gasparrini}, {Giglietto}, {Giordano}, {Giroletti}, {Green}, {Grenier},
  {Grove}, {Guiriec}, {Hays}, {Hewitt}, {Horan}, {J{\'o}hannesson}, {Kocevski},
  {Kuss}, {Latronico}, {Li}, {Longo}, {Loparco}, {Lovellette}, {Lubrano},
  {Maldera}, {Manfreda}, {Mart{\'\i}-Devesa}, {Mazziotta}, {Mereu}, {Meyer},
  {Michelson}, {Mirabal}, {Mitthumsiri}, {Mizuno}, {Monzani}, {Moretti},
  {Morselli}, {Moskalenko}, {Negro}, {Nuss}, {Ohno}, {Omodei}, {Orienti},
  {Orlando}, {Palatiello}, {Paliya}, {Paneque}, {Persic}, {Pesce-Rollins},
  {Petrosian}, {Piron}, {Poolakkil}, {Poon}, {Porter}, {Principe}, {Racusin},
  {Rain{\`o}}, {Rando}, {Razzano}, {Razzaque}, {Reimer}, {Reimer}, {Reposeur},
  {Ryde}, {Serini}, {Sgr{\`o}}, {Siskind}, {Sonbas}, {Spandre}, {Spinelli},
  {Suson}, {Tajima}, {Takahashi}, {Tak}, {Thayer}, {Torres}, {Troja},
  {Valverde}, {Veres}, {Vianello}, {von Kienlin}, {Wood}, {Yassine}, {Zhu}, \&
  {Zimmer}}]{Ajello_etal_2019}
{Ajello}, M., {Arimoto}, M., {Axelsson}, M., {et~al.} 2019, \apj, 878, 52,
  \dodoi{10.3847/1538-4357/ab1d4e}

\bibitem[{Alexander {et~al.}(2017)Alexander, Laskar, Berger, Guidorzi,
  Dichiara, Fong, Gomboc, Kobayashi, Kopac, Mundell, {et~al.}}]{alexander2017}
Alexander, K.~D., Laskar, T., Berger, E., {et~al.} 2017, The Astrophysical
  Journal, 848, 69

\bibitem[{{Barniol Duran} \& {Kumar}(2009)}]{Barniol_Kumar2009}
{Barniol Duran}, R., \& {Kumar}, P. 2009, \mnras, 395, 955,
  \dodoi{10.1111/j.1365-2966.2009.14584.x}

\bibitem[{{Beniamini} {et~al.}(2017){Beniamini}, {Giannios}, \&
  {Metzger}}]{Beniamini_etal_2017}
{Beniamini}, P., {Giannios}, D., \& {Metzger}, B.~D. 2017, \mnras, 472, 3058,
  \dodoi{10.1093/mnras/stx2095}

\bibitem[{{Bernardini}(2015)}]{Bernardini_2015_magnetar}
{Bernardini}, M.~G. 2015, Journal of High Energy Astrophysics, 7, 64,
  \dodoi{10.1016/j.jheap.2015.05.003}

\bibitem[{{Blandford} \& {Znajek}(1977)}]{Blandford_Znajek_1977}
{Blandford}, R.~D., \& {Znajek}, R.~L. 1977, \mnras, 179, 433,
  \dodoi{10.1093/mnras/179.3.433}

\bibitem[{Bloom {et~al.}(2001)Bloom, Frail, \& Sari}]{bloom2001prompt}
Bloom, J.~S., Frail, D.~A., \& Sari, R. 2001, The Astronomical Journal, 121,
  2879

\bibitem[{Cenko {et~al.}(2011)Cenko, Frail, Harrison, Haislip, Reichart,
  Butler, Cobb, Cucchiara, Berger, Bloom, {et~al.}}]{cenko2011}
Cenko, S., Frail, D., Harrison, F., {et~al.} 2011, The Astrophysical Journal,
  732, 29

\bibitem[{{Chakrabarty}(2008)}]{Chakrabarty2008}
{Chakrabarty}, D. 2008, in American Institute of Physics Conference Series,
  Vol. 1068, American Institute of Physics Conference Series, ed.
  R.~{Wijnands}, D.~{Altamirano}, P.~{Soleri}, N.~{Degenaar}, N.~{Rea},
  P.~{Casella}, A.~{Patruno}, \& M.~{Linares}, 67--74,
  \dodoi{10.1063/1.3031208}

\bibitem[{Chand {et~al.}(2019)Chand, Chattopadhyay, Oganesyan, Rao, Vadawale,
  Bhattacharya, Bhalerao, \& Misra}]{chand2019}
Chand, V., Chattopadhyay, T., Oganesyan, G., {et~al.} 2019, The Astrophysical
  Journal, 874, 70

\bibitem[{{Chen} {et~al.}(2017){Chen}, {Xie}, {Lei}, {Zou}, {L{\"u}}, {Liang},
  {Gao}, \& {Wang}}]{Chen_etal_2017_magnetar_to_BH}
{Chen}, W., {Xie}, W., {Lei}, W.-H., {et~al.} 2017, \apj, 849, 119,
  \dodoi{10.3847/1538-4357/aa8f4a}

\bibitem[{Chen {et~al.}(2020)Chen, Urata, Huang, Takahashi, Petitpas, \&
  Asada}]{chen2020}
Chen, W.~J., Urata, Y., Huang, K., {et~al.} 2020, arXiv preprint
  arXiv:2002.06722

\bibitem[{{Chen} \& {Beloborodov}(2007)}]{Chen_Beloborodov_2007}
{Chen}, W.-X., \& {Beloborodov}, A.~M. 2007, \apj, 657, 383,
  \dodoi{10.1086/508923}

\bibitem[{{Cromartie} {et~al.}(2020){Cromartie}, {Fonseca}, {Ransom},
  {Demorest}, {Arzoumanian}, {Blumer}, {Brook}, {DeCesar}, {Dolch}, {Ellis},
  {Ferdman}, {Ferrara}, {Garver-Daniels}, {Gentile}, {Jones}, {Lam}, {Lorimer},
  {Lynch}, {McLaughlin}, {Ng}, {Nice}, {Pennucci}, {Spiewak}, {Stairs},
  {Stovall}, {Swiggum}, \& {Zhu}}]{Cromartie_etal_2020_maxNSmass}
{Cromartie}, H.~T., {Fonseca}, E., {Ransom}, S.~M., {et~al.} 2020, Nature
  Astronomy, 4, 72, \dodoi{10.1038/s41550-019-0880-2}

\bibitem[{{Cunningham} {et~al.}(2020){Cunningham}, {Cenko}, {Ryan}, {Vogel},
  {Corsi}, {Cucchiara}, {Fruchter}, {Horesh}, {Kangas}, {Kocevski}, {Perley},
  \& {Racusin}}]{Cunningham_etal_2020}
{Cunningham}, V., {Cenko}, S.~B., {Ryan}, G., {et~al.} 2020, arXiv e-prints,
  arXiv:2009.00579.
\newblock \doarXiv{2009.00579}

\bibitem[{{Duncan} \& {Thompson}(1992)}]{Duncan_Thompson1992}
{Duncan}, R.~C., \& {Thompson}, C. 1992, \apjl, 392, L9, \dodoi{10.1086/186413}

\bibitem[{Evans {et~al.}(2007)Evans, Beardmore, Page, Tyler, Osborne, Goad,
  O'Brien, Vetere, Racusin, Morris, {et~al.}}]{evans2007online}
Evans, P., Beardmore, A., Page, K.~L., {et~al.} 2007, Astronomy \&
  Astrophysics, 469, 379

\bibitem[{Evans {et~al.}(2009)Evans, Beardmore, Page, Osborne, O'Brien,
  Willingale, Starling, Burrows, Godet, Vetere, {et~al.}}]{evans2009methods}
Evans, P., Beardmore, A., Page, K., {et~al.} 2009, Monthly Notices of the Royal
  Astronomical Society, 397, 1177

\bibitem[{Frail {et~al.}(2001)Frail, Kulkarni, Sari, Djorgovski, Bloom, Galama,
  Reichart, Berger, Harrison, Price, {et~al.}}]{frail2001beaming}
Frail, D.~A., Kulkarni, S., Sari, R., {et~al.} 2001, The Astrophysical Journal
  Letters, 562, L55

\bibitem[{Goldstein {et~al.}(2016)Goldstein, Connaughton, Briggs, \&
  Burns}]{goldstein2016estimating}
Goldstein, A., Connaughton, V., Briggs, M.~S., \& Burns, E. 2016, The
  Astrophysical Journal, 818, 18

\bibitem[{Granot {et~al.}(2002)Granot, Panaitescu, Kumar, \&
  Woosley}]{granot2002off}
Granot, J., Panaitescu, A., Kumar, P., \& Woosley, S.~E. 2002, The
  Astrophysical Journal Letters, 570, L61

\bibitem[{{Hessels} {et~al.}(2006){Hessels}, {Ransom}, {Stairs}, {Freire},
  {Kaspi}, \& {Camilo}}]{fastest_pulsar2006}
{Hessels}, J. W.~T., {Ransom}, S.~M., {Stairs}, I.~H., {et~al.} 2006, Science,
  311, 1901, \dodoi{10.1126/science.1123430}

\bibitem[{{Kumar} {et~al.}(2008){Kumar}, {Narayan}, \&
  {Johnson}}]{Pawan_etal_2008}
{Kumar}, P., {Narayan}, R., \& {Johnson}, J.~L. 2008, \mnras, 388, 1729,
  \dodoi{10.1111/j.1365-2966.2008.13493.x}

\bibitem[{Laskar {et~al.}(2016)Laskar, Alexander, Berger, Fong, Margutti,
  Shivvers, Williams, Kopac, Kobayashi, Mundell, {et~al.}}]{laskar2016}
Laskar, T., Alexander, K.~D., Berger, E., {et~al.} 2016, arXiv preprint
  arXiv:1606.08873

\bibitem[{{Lee} {et~al.}(2000){Lee}, {Wijers}, \& {Brown}}]{Lee_etal_2000}
{Lee}, H.~K., {Wijers}, R.~A.~M.~J., \& {Brown}, G.~E. 2000, \physrep, 325, 83,
  \dodoi{10.1016/S0370-1573(99)00084-8}

\bibitem[{{Lei} {et~al.}(2017){Lei}, {Zhang}, {Wu}, \& {Liang}}]{Lei_etal_2017}
{Lei}, W.-H., {Zhang}, B., {Wu}, X.-F., \& {Liang}, E.-W. 2017, \apj, 849, 47,
  \dodoi{10.3847/1538-4357/aa9074}

\bibitem[{{Leng} \& {Giannios}(2014)}]{Leng_Giannios_2014}
{Leng}, M., \& {Giannios}, D. 2014, \mnras, 445, L1,
  \dodoi{10.1093/mnrasl/slu122}

\bibitem[{{Li} {et~al.}(2018){Li}, {Wu}, {Lei}, {Dai}, {Liang}, \&
  {Ryde}}]{Liang_etal_2018}
{Li}, L., {Wu}, X.-F., {Lei}, W.-H., {et~al.} 2018, \apjs, 236, 26,
  \dodoi{10.3847/1538-4365/aabaf3}

\bibitem[{Li {et~al.}(2018)Li, Wu, Lei, Dai, Liang, \& Ryde}]{li2018}
Li, L., Wu, X.-F., Lei, W.-H., {et~al.} 2018, The Astrophysical Journal
  Supplement Series, 236, 26

\bibitem[{{MacFadyen} \& {Woosley}(1999)}]{MacFadyen_Woosley_1999}
{MacFadyen}, A.~I., \& {Woosley}, S.~E. 1999, \apj, 524, 262,
  \dodoi{10.1086/307790}

\bibitem[{{MacLachlan} {et~al.}(2013){MacLachlan}, {Shenoy}, {Sonbas}, {Dhuga},
  {Cobb}, {Ukwatta}, {Morris}, {Eskandarian}, {Maximon}, \&
  {Parke}}]{MacLachlan_etal_2013}
{MacLachlan}, G.~A., {Shenoy}, A., {Sonbas}, E., {et~al.} 2013, \mnras, 432,
  857, \dodoi{10.1093/mnras/stt241}

\bibitem[{McKinney(2005)}]{mckinney2005jet}
McKinney, J.~C. 2005, arXiv preprint astro-ph/0506369

\bibitem[{Meszaros(2006)}]{meszaros2006gamma}
Meszaros, P. 2006, Reports on Progress in Physics, 69, 2259

\bibitem[{Metzger {et~al.}(2011)Metzger, Giannios, Thompson, Bucciantini, \&
  Quataert}]{Metzger_etal_2011}
Metzger, B., Giannios, D., Thompson, T., Bucciantini, N., \& Quataert, E. 2011,
  Monthly Notices of the Royal Astronomical Society, 413, 2031

\bibitem[{{Metzger}(2017)}]{Metzger2017}
{Metzger}, B.~D. 2017, Living Reviews in Relativity, 20, 3,
  \dodoi{10.1007/s41114-017-0006-z}

\bibitem[{{Metzger} {et~al.}(2018){Metzger}, {Beniamini}, \&
  {Giannios}}]{Metzger_etal_2018}
{Metzger}, B.~D., {Beniamini}, P., \& {Giannios}, D. 2018, \apj, 857, 95,
  \dodoi{10.3847/1538-4357/aab70c}

\bibitem[{Misra {et~al.}(2019)Misra, Resmi, Kann, Marongiu, Moin, Klose,
  Postigo, Jaiswal, Perley, Ghosh, {et~al.}}]{misra2019}
Misra, K., Resmi, L., Kann, D., {et~al.} 2019, arXiv preprint arXiv:1911.09719

\bibitem[{{Narayan} {et~al.}(1992){Narayan}, {Paczynski}, \&
  {Piran}}]{Narayan_etal_1992}
{Narayan}, R., {Paczynski}, B., \& {Piran}, T. 1992, \apjl, 395, L83,
  \dodoi{10.1086/186493}

\bibitem[{Narayan {et~al.}(2001)Narayan, Piran, \&
  Kumar}]{narayan2001accretion}
Narayan, R., Piran, T., \& Kumar, P. 2001, The Astrophysical Journal, 557, 949

\bibitem[{{Nathanail} {et~al.}(2016){Nathanail}, {Strantzalis}, \&
  {Contopoulos}}]{Nathanail_etal_2016}
{Nathanail}, A., {Strantzalis}, A., \& {Contopoulos}, I. 2016, \mnras, 455,
  4479, \dodoi{10.1093/mnras/stv2558}

\bibitem[{Nemmen {et~al.}(2012)Nemmen, Georganopoulos, Guiriec, Meyer, Gehrels,
  \& Sambruna}]{nemmen2012}
Nemmen, R.~S., Georganopoulos, M., Guiriec, S., {et~al.} 2012, Science, 338,
  1445

\bibitem[{{{\"O}zel} {et~al.}(2012){{\"O}zel}, {Psaltis}, {Narayan}, \& {Santos
  Villarreal}}]{Ozel_etal_2012}
{{\"O}zel}, F., {Psaltis}, D., {Narayan}, R., \& {Santos Villarreal}, A. 2012,
  \apj, 757, 55, \dodoi{10.1088/0004-637X/757/1/55}

\bibitem[{{Papitto} {et~al.}(2014){Papitto}, {Torres}, {Rea}, \&
  {Tauris}}]{Papitto_etal_2014}
{Papitto}, A., {Torres}, D.~F., {Rea}, N., \& {Tauris}, T.~M. 2014, \aap, 566,
  A64, \dodoi{10.1051/0004-6361/201321724}

\bibitem[{{Patruno} {et~al.}(2017){Patruno}, {Haskell}, \&
  {Andersson}}]{Patruno_etal_2017}
{Patruno}, A., {Haskell}, B., \& {Andersson}, N. 2017, \apj, 850, 106,
  \dodoi{10.3847/1538-4357/aa927a}

\bibitem[{{Peng} {et~al.}(2005){Peng}, {K{\"o}nigl}, \&
  {Granot}}]{Peng_etal_2005}
{Peng}, F., {K{\"o}nigl}, A., \& {Granot}, J. 2005, \apj, 626, 966,
  \dodoi{10.1086/430045}

\bibitem[{{Perley} {et~al.}(2014){Perley}, {Cenko}, {Corsi}, {Tanvir}, {Levan},
  {Kann}, {Sonbas}, {Wiersema}, {Zheng}, {Zhao}, {Bai}, {Bremer},
  {Castro-Tirado}, {Chang}, {Clubb}, {Frail}, {Fruchter},
  {G{\"o}{\u{g}}{\"u}{\textcommabelow s}}, {Greiner}, {G{\"u}ver}, {Horesh},
  {Filippenko}, {Klose}, {Mao}, {Morgan}, {Pozanenko}, {Schmidl}, {Stecklum},
  {Tanga}, {Volnova}, {Volvach}, {Wang}, {Winters}, \& {Xin}}]{perley2014}
{Perley}, D.~A., {Cenko}, S.~B., {Corsi}, A., {et~al.} 2014, \apj, 781, 37,
  \dodoi{10.1088/0004-637X/781/1/37}

\bibitem[{Pisani {et~al.}(2016)Pisani, Ruffini, Aimuratov, Bianco, Kovacevic,
  Moradi, Muccino, Penacchioni, Rueda, Shakeri, {et~al.}}]{pisani2016}
Pisani, G., Ruffini, R., Aimuratov, Y., {et~al.} 2016, arXiv preprint
  arXiv:1610.05619

\bibitem[{{Racusin} {et~al.}(2011){Racusin}, {Oates}, {Schady}, {Burrows}, {de
  Pasquale}, {Donato}, {Gehrels}, {Koch}, {McEnery}, {Piran}, {Roming},
  {Sakamoto}, {Swenson}, {Troja}, {Vasileiou}, {Virgili}, {Wanderman}, \&
  {Zhang}}]{Racusin_etal_2011}
{Racusin}, J.~L., {Oates}, S.~R., {Schady}, P., {et~al.} 2011, \apj, 738, 138,
  \dodoi{10.1088/0004-637X/738/2/138}

\bibitem[{{Rowlinson} {et~al.}(2014){Rowlinson}, {Gompertz}, {Dainotti},
  {O'Brien}, {Wijers}, \& {van der Horst}}]{Rowlinson_etal_2014}
{Rowlinson}, A., {Gompertz}, B.~P., {Dainotti}, M., {et~al.} 2014, \mnras, 443,
  1779, \dodoi{10.1093/mnras/stu1277}

\bibitem[{{Ruffert} {et~al.}(1997){Ruffert}, {Janka}, {Takahashi}, \&
  {Schaefer}}]{Ruffert_etal_1997}
{Ruffert}, M., {Janka}, H.~T., {Takahashi}, K., \& {Schaefer}, G. 1997, \aap,
  319, 122.
\newblock \doarXiv{astro-ph/9606181}

\bibitem[{{Ryan} {et~al.}(2015){Ryan}, {van Eerten}, {MacFadyen}, \&
  {Zhang}}]{Ryan_etal_2015}
{Ryan}, G., {van Eerten}, H., {MacFadyen}, A., \& {Zhang}, B.-B. 2015, \apj,
  799, 3, \dodoi{10.1088/0004-637X/799/1/3}

\bibitem[{Sari {et~al.}(1999)Sari, Piran, \& Halpern}]{sari1999_jets}
Sari, R., Piran, T., \& Halpern, J. 1999, arXiv preprint astro-ph/9903339

\bibitem[{{Sarin} {et~al.}(2019){Sarin}, {Lasky}, \&
  {Ashton}}]{Sarin_etal_2019}
{Sarin}, N., {Lasky}, P.~D., \& {Ashton}, G. 2019, \apj, 872, 114,
  \dodoi{10.3847/1538-4357/aaf9a0}

\bibitem[{{Shapiro} \& {Shibata}(2002)}]{Shapiro_Shibata_2002}
{Shapiro}, S.~L., \& {Shibata}, M. 2002, \apj, 577, 904, \dodoi{10.1086/342246}

\bibitem[{{Shibata} \& {Shapiro}(2002)}]{Shibata_Shapiro_2002}
{Shibata}, M., \& {Shapiro}, S.~L. 2002, \apjl, 572, L39,
  \dodoi{10.1086/341516}

\bibitem[{Song \& Liu(2019)}]{song2019}
Song, C.-Y., \& Liu, T. 2019, The Astrophysical Journal, 871, 117

\bibitem[{Song {et~al.}(2016)Song, Liu, Gu, \& Tian}]{song2016}
Song, C.-Y., Liu, T., Gu, W.-M., \& Tian, J.-X. 2016, Monthly Notices of the
  Royal Astronomical Society, 458, 1921

\bibitem[{Song {et~al.}(2018)Song, Liu, \& Li}]{song2018}
Song, C.-Y., Liu, T., \& Li, A. 2018, Monthly Notices of the Royal Astronomical
  Society, 477, 2173

\bibitem[{Tachibana {et~al.}(2018)Tachibana, Arimoto, Asano, Harita, Fujiwara,
  Yoshii, Itoh, Murata, Yatsu, Morita, {et~al.}}]{tachibana2018}
Tachibana, Y., Arimoto, M., Asano, K., {et~al.} 2018, Publications of the
  Astronomical Society of Japan, 70, 92

\bibitem[{{Usov}(1992)}]{Usov1992}
{Usov}, V.~V. 1992, \nat, 357, 472, \dodoi{10.1038/357472a0}

\bibitem[{{van Eerten} {et~al.}(2010){van Eerten}, {Zhang}, \&
  {MacFadyen}}]{VanEerten_etal_2010}
{van Eerten}, H., {Zhang}, W., \& {MacFadyen}, A. 2010, \apj, 722, 235,
  \dodoi{10.1088/0004-637X/722/1/235}

\bibitem[{Wang {et~al.}(2018)Wang, Zhang, Liang, Lu, Lin, Li, \& Li}]{wang2018}
Wang, X.-G., Zhang, B., Liang, E.-W., {et~al.} 2018, The Astrophysical Journal,
  859, 160

\bibitem[{Wang {et~al.}(2015)Wang, Zhang, Liang, Gao, Li, Deng, Qin, Tang,
  Kann, Ryde, {et~al.}}]{wang2015_break}
---. 2015, The Astrophysical Journal Supplement Series, 219, 9

\bibitem[{{Wiktorowicz} {et~al.}(2014){Wiktorowicz}, {Belczynski}, \&
  {Maccarone}}]{Wiktorowicz_etal_2014}
{Wiktorowicz}, G., {Belczynski}, K., \& {Maccarone}, T. 2014, in Binary
  Systems, their Evolution and Environments, 37.
\newblock \doarXiv{1312.5924}

\bibitem[{Woosley(1993)}]{woosley1993gamma}
Woosley, S. 1993, The Astrophysical Journal, 405, 273

\bibitem[{{Yamazaki} {et~al.}(2020){Yamazaki}, {Sato}, {Sakamoto}, \&
  {Serino}}]{Yamazaki_etal_2020}
{Yamazaki}, R., {Sato}, Y., {Sakamoto}, T., \& {Serino}, M. 2020, \mnras, 494,
  5259, \dodoi{10.1093/mnras/staa1095}

\bibitem[{Zhang {et~al.}(2015)Zhang, Van~Eerten, Burrows, Ryan, Evans, Racusin,
  Troja, \& MacFadyen}]{zhang2015}
Zhang, B.-B., Van~Eerten, H., Burrows, D.~N., {et~al.} 2015, The Astrophysical
  Journal, 806, 15

\bibitem[{{Zhao} {et~al.}(2020){Zhao}, {Liu}, {Gao}, {Lan}, {Lei}, \&
  {Xie}}]{Zhao_etal_2020}
{Zhao}, L., {Liu}, L., {Gao}, H., {et~al.} 2020, \apj, 896, 42,
  \dodoi{10.3847/1538-4357/ab8f91}

\end{thebibliography}
\bibliographystyle{aasjournal}



\end{document}